# Directional intermodular coupling enriches functional complexity in biological neuronal networks

Nobuaki Monma[1,2], Hideaki Yamamoto[1,2]*, Naoya Fujiwara[3,4], Hakuba Murota[1,5], Satoshi Moriya[1], Ayumi Hirano-Iwata[1,2,5,6], and Shigeo Sato[1,2]

[1]*Research Institute of Electrical Communication, Tohoku University, Sendai, Japan*
[2]*Graduate School of Engineering, Tohoku University, Sendai, Japan*
[3]*Graduate School of Information Sciences, Tohoku University, Sendai, Japan*
[4]*International Research Center for Neurointelligence (WPI-IRCN), UTIAS, The University of Tokyo, Tokyo, Japan*
[5]*Graduate School of Biomedical Engineering, Tohoku University, Sendai, Japan*
[6]*WPI-Advanced Institute for Materials Research, Tohoku University, Sendai, Japan*

***Corresponding author at:** Research Institute of Electrical Communication, Tohoku University, Sendai 980-8577, Japan
E-mail address: hideaki.yamamoto.e3@tohoku.ac.jp (Hideaki Yamamoto)

**Highlights**

- Microfluidic devices realize hierarchically modular networks with directionality.
- Modularity and directionality corporate to increase dynamical complexity.
- Bioengineering technologies link wet experiments to theoretical models.

***Abstract***
Hierarchically modular organization is a canonical network topology that is evolutionarily conserved in the nervous systems of animals. Within the network, neurons form directional connections defined by the growth of their axonal terminals. However, this topology is dissimilar to the network formed by dissociated neurons in culture because they form randomly connected networks on homogeneous substrates. In this study, we fabricated microfluidic devices to reconstitute hierarchically modular neuronal networks in culture (in vitro) and investigated how non-random structures, such as directional connectivity between modules, affect global network dynamics. Embedding directional connections in a pseudo-feedforward manner suppressed excessive synchrony in cultured neuronal networks and enhanced the integration-segregation balance. Modeling the behavior of biological neuronal networks using spiking neural networks (SNNs) further revealed that modularity and directionality cooperate to shape such network dynamics. Finally, we demonstrate that for a given network topology, the statistics of network dynamics, such as global network activation, correlation coefficient, and functional complexity, can be analytically predicted based on eigendecomposition of the transition matrix in the state-transition model. Hence, the integration of bioengineering and cell culture technologies enables us not only to reconstitute complex network circuitry in the nervous system but also to understand the structure-function relationships in biological neuronal networks by bridging theoretical modeling with in vitro experiments.

**Keywords: Neuronal network, axon, bioengineering, spiking neural network, computational neuroscience**

## *1. Introduction*

Brain-on-a-chip technology has emerged as a tool for modeling the complex connectivity and functions of neuronal networks in the brain with living cells. Such tools not only provide a novel platform to effectively screen candidate compounds in drug discovery (Ma et al., 2015) but also aid fundamental research in molecular, cellular, and network neuroscience (Lassus et al., 2018; Pigareva et al., 2023; Sakaibara et al., 2024). Among the various levels of biological hierarchy, network-level studies have especially profited, as technology offers a method to reconstitute neuronal networks with a well-defined structure and unravels their correlation to emergent activity patterns (Yamamoto et al., 2018a; Park et al., 2021; Montalà-Flaquer et al., 2022).

Directional connectivity is one of the most fundamental nonrandom properties of brain networks, as neurons are polarized cells with axonal and dendritic fibers that are responsible for sending outputs and receiving inputs, respectively. In the local circuits of the mammalian cortex, for example, unidirectional (nonreciprocal) connections are more abundant than bidirectional (reciprocal) connections (Markram et al., 1997; Song et al., 2005). In cultured neuronal networks in vitro, a top-down control of directionality has been demonstrated using asymmetric microchannels (Peyrin et al., 2011; Renault et al., 2016; Gladkov et al., 2017; Holloway et al., 2019; Pigareva et al., 2021; Courte et al., 2023; Winter-Hjelm et al., 2023), triangular micropatterns (Feinerman et al., 2008; Isomura et al., 2015; Albers & Offenhausser, 2016), stomach-shaped micropatterns (Forró et al., 2018; Girardin et al., 2022), or asymmetric cross-shaped micropatterns (Yamamoto et al., 2016; Ming et al., 2021). The fundamental idea behind these experiments is that without any guidance cues, neurons in dissociated cultures tend to grow neurites in random orientations and form overabundant reciprocal connections (Stetter et al., 2012).

Another canonical structure that is evolutionarily conserved in the animal nervous system is the modular structure, in which a network is composed of multiple densely coupled subnetworks that interact sparsely. Previous studies using micropatterned substrates and microfluidic devices revealed that such a network structure helps the network suppress excessive network synchrony in cultured neuronal networks to balance integrated (synchronous firing) states with segregated (asynchronous firing) states (Yamamoto et al., 2018a; Takemuro et al., 2020; Sato et al., 2023), a critical portrait of brain dynamics. Reconstitution of directional connectivity and modular organization are not mutually exclusive (Forró et al. 2018; Girardin et al., 2022). Indeed, reduction of direct reciprocal connections between modules are hypothesized to suppress the excessive synchrony in the cultured neuronal network, since synaptic transmission creates a temporal delay of ~0.6 ms (Lisman et al., 2007) and signal propagation velocity along an axon is finite at ~1 m/s (Bakkum et al, 2013). However, a quantitative understanding of how directional connections affect global network activity is lacking, as is an experimental platform to verify theoretical predictions.

In this study, we constructed cultured (in vitro) neuronal networks with well-defined connectivity and revealed that the introduction of directional connections is effective in reducing excessive synchrony and balancing the integrated and segregated states. Next, we numerically investigated in silico how the selectivity of directional connections embedded in inter-modular couplings impacts the network synchrony using spiking neural networks (SNNs). Finally, we built a simple state-transition model to analytically formulate the structure-functional correlation. The work reveals functional significance of directional intermodular connections in neuronal networks with directional inter-modular connections and, simultaneously, advances the field of brain-on-a-chip technology in reconstituting in vitro systems that more-closely resemble the structure and function of the brain networks.

## 2. *Materials & Methods*

### *2.1. In vitro biological experiments*

#### *2.1.1. Microfluidic devices*

Microfluidic devices were fabricated using polydimethylsiloxane (PDMS) via the replica molding method (Takemuro et al., 2020). Briefly, a negative master mold was first fabricated by patterning two layers of photoresist, SU-8 3005 and SU-8 3050, on a 2-inch silicon wafer. The three-dimensional structures of the master mold and microfluidic film were evaluated using confocal microscopy (Keyence VK-X260). Subsequently, PDMS gel (Sylgard 184, base agent: curing agent = 7.5:1 (w/w)) was poured onto the master mold, cured in an oven for 3 h at 60 °C, and then peeled off from the mold with forceps. Prior to cell culture, the microfluidic film was sterilized by UV irradiation and subsequently attached to a glass coverslip pre-coated with poly-D-lysine (PDL).

#### *2.1.2. Cell culture*

The microfluidic device attached to the PDL-coated coverslip was first immersed in a neuronal plating medium [Minimum Essential Medium (MEM) (Gibco 11095-080) + 5% fetal bovine serum + 0.6% D-glucose)] and stored overnight in an incubator. Rat cortical neurons were obtained from E18 embryos and cultured according to previously published protocols (Yamamoto et al., 2018a). Briefly, the dissociated cells were suspended in plating medium at a concentration of 7.5–10 $\times 10^5$ cells/cm$^2$. The coverslip with the microfluidic device was then transferred and placed upside down in a cell culture dish with glial cells grown in N2 medium [MEM + 10% N2 supplement + 1% ovalbumin (Sigma 17504-044) + 1M HEPES]. At 5 days in vitro (DIV), cytosine arabinoside (1 μM) was added to prevent glial cell proliferation. All animal experiments were approved by the Tohoku University Center for Laboratory Animal Research, Tohoku University (approval number: 2020AmA-001).

### *2.1.3. Neurite elongation*

Neurons were seeded at a lower concentration ($0.5\times10^4$ cells), and time-lapse images were obtained for 40 h at 4 frames per hour starting at 1 day in vitro (DIV). The imaging was performed using an inverted microscope (IX83, Olympus) equipped with a 40× objective lens (numerical aperture, 0.95), a scientific CMOS camera (ORCA-Fusion BT, Hamamatsu Photonics), and a stage top incubator (TOKAI HIT; 37 °C, 5% $CO_2$). The microscope was controlled using CellSens software (Olympus). The selectivity of neurite penetration through the microchannels was assessed by counting the number of neurites that penetrated the microchannels in the forward direction and dividing the value by the total number of neurites that penetrated the microchannels in either direction.

### *2.1.4. Calcium imaging*

Cultured neurons were transfected with the fluorescent calcium probe GCaMP6s using an adeno-associated virus (AAV) vector. At 4 DIV, half of the medium was removed from the petri dish, AAV was added at a concentration of 0.2 ng/μL, and the previously removed medium was reintroduced on the next day. At 10 DIV, the cells were transferred to the HEPES-buffered saline (128 mM NaCl, 4 mM KCl, 1 mM $CaCl_2$, 1mM $MgCl_2$, 10 mM D-glucose, 10 mM HEPES, and 45 mM sucrose), and fluorescence calcium imaging was performed using an inverted microscope (IX83, Olympus) equipped with a 20× objective lens (numerical aperture, 0.70), a white light LED (X-Cite XYLIS, Excelitas), scientific CMOS camera (ORCA-Fusion BT, Hamamatsu Photonics), and stage top incubator (37 °C). Three or four networks were selected from a coverslip, and spontaneous neural activity was recorded for 20 min at 100 frames per second (fps) using HCImage software (Hamamatsu Photonics). All experiments involving viral vectors were approved by the Tohoku University Center for Gene Research (2021EeLMO-001).

### *2.1.5. Data acquisition*

To obtain neuronal activity from calcium imaging recordings, approximately 64 neurons (four neurons per module) were selected from a single network, and a region of interest (ROI) was drawn surrounding each neuron using Fiji (National Institute of Health). The fluorescent value $F(t)$ was obtained by averaging all pixel values within the ROI at timestep $t$. To remove the fading effects and background noise in $F(t)$, the relative fluorescence $\Delta F(t)/F_0(t)$ was calculated, given by $(F(t){-}F_0(t))/F_0(t)$, where $F_0(t)$ is the baseline of the ROI.

To infer the firing rate of neuron $i$, $x_i(\mathrm{t})$, from the relative fluorescence $\Delta F(t)/F_0(t)$, we employed the CASCADE algorithm, which is a convolutional neural network trained on data obtained in simultaneous fluorescence calcium imaging and patch clamp experiments (Rupprecht et al., 2021). We re-trained the model to match our experimental conditions (100 fps, GCaMP6s, Gaussian kernel with a standard deviation (SD) of 25 ms) and used it for inference.

### *2.1.6. Signal Propagation*

To evaluate the signal propagation between modules, first, the activation onset of neuron $i$ was detected by adapting the Schmitt trigger algorithm to $x_i(t)$. Then, a binary time series $S_i(t)$ was created, where a value of 1 is associated with the detection of an onset (upper threshold: 2 Hz, lower threshold: 1 Hz). Subsequently, a collective activity event was detected by adapting Schmitt trigger method to population firing rate obtained by summing $x_i(t)$ across all neurons (upper threshold: 10 Hz, lower threshold: 5 Hz). When the detected event involved only a single neuron, the event was excluded from further analysis. The onset time of the collective activity event was then redefined to be 0.2 s earlier than the detected onset time to include neuronal spikes that occurred before the population firing rate crossed the upper threshold. Finally, within the event, the propagation probability and propagation delay were quantified based on $S_i(t)$ of the neurons affiliated to the modules.

In addition, a cross-correlation function was used to evaluate spike statistics, especially the effect of the number of asymmetric microchannels that connect a pair of modules on their correlated activity. First, the cross-correlation function for a given pair of neurons $i$ and $j$ was given by:

$$R_{ij}(\tau) = \frac{1}{T-\tau} \sum_{l=0}^{(T-\tau)/\Delta t - 1} S_i(l\Delta t)\, S_j(l\Delta t - \tau)\,, \tag{1}$$

where $T$ is the duration of a recording (= 1200 s), $\Delta t$ is the time step (= 0.01 s), and $\tau$ is the time delay ($-1.0 < \tau < 1.0$ s). The indices $i$ and $j$ are assigned to neurons in the upstream and downstream modules, to set negative and positive delays for reverse and forward propagation between modules, respectively. Then, accumulated cross-correlation function $\langle \hat{R}_{ij}(\tau) \rangle_h$ was obtained by first normalizing the cross-correlation function as $\hat{R}_{ij}(\tau) = \frac{R_{ij}(\tau)}{\sqrt{R_{ii}(0) R_{jj}(0)}}$ and then averaging it over all $i$-$j$ pairs with a hop of $h$, which was defined as the minimum number of asymmetric microchannels required to reach from neuron $i$ to neuron $j$ in the forward direction.

### *2.1.7. Analysis of network dynamics*

Measurements such as global network activation (GNA), mean correlation, and functional complexity (Zamora-López et al., 2016) were computed to assess network dynamics. GNA (Φ) was defined as the average fraction of neurons involved in collective activity events. The Pearson's correlation coefficient between neurons $i$ and $j$, $r_{ij}$, was computed as follows:

$$r_{ij} = \frac{\sum_t [x_i(t) - \bar{x}_i] \cdot [x_j(t) - \bar{x}_j]}{\sqrt{\sum_t [x_i(t) - \bar{x}_i]^2} \sqrt{\sum_t [x_j(t) - \bar{x}_j]^2}}, \tag{2}$$

where $x_i(t)$ is the firing rate of neuron $i$ at time $t$, and $\bar{x}_i$ is the time average of $x_i(t)$. The mean correlation coefficient was calculated as $\langle r_{ij} \rangle = \frac{1}{M} \sum_{i \neq j} r_{ij} (1 - \delta(m_i, m_j))$, where $m_i$ is index of the modules to which neuron $i$ belongs, $\delta(\cdot,\cdot)$ is the Kronecker delta, and $M = \sum_{i \neq j} \left(1 - \delta\left(m_i, m_j\right)\right)$ is the normalization constant. Subsequently, the functional complexity is computed from the correlation matrix $r_{ij}$, given by:

$$\Theta = 1 - \frac{B}{2(B-1)} \sum_{k=1}^{B} \left| p_k\left(r_{ij}\right) - \frac{1}{B} \right|, \tag{3}$$

where $p_k(r_{ij})$ is the probability distribution of the correlation coefficients at $k$-th bin, and $B$ is the number of histogram bins (= 20). Since the correlation coefficient between neurons within the same module is high (with $r_{ij}$ almost equal to 1), $p_k(r_{ij})$ was calculated by focusing only on the correlation coefficient corresponding the $i$-$j$ neuron pairs in different modules. When the value of the correlation coefficient is localized at approximately 0 or 1, functional complexity decreases. Conversely, when the value of the correlation coefficients is widely distributed, the functional complexity increases. Thus, this measure assesses the balance between integration and segregation states.

Only the networks consisting of 110–160 neurons were used for the analysis to eliminate the influence of cell number. The number of neurons in the module was manually determined from phase-contrast images (mean ± SD, non-directional networks, 130 ± 11 cells ($n$ = 12); directional networks, 135 ± 16 cells ($n$ = 10); $p > 0.05$, Student's $t$-test). If the cell population aggregated in the module, the number of cells was inferred from its area, assuming that a single cell occupies 60 pixels. This value was the median number of pixels for a cell body in homogeneous cultures ($n$ = 282). Moreover, neurons selected as ROIs but exhibited less than two activity onsets were regarded as non-active neurons and were ignored in the analysis.

### *2.2. Spiking neural network*

#### *2.2.1. Neuron model*

Each neuron in the SNN model was modeled using the Izhikevich model (Izhikevich, 2003). The model is based on a two-variable system of differential equations that describe the membrane potential $v$ and recovery variable $u$ of neuron $i$ as:

$$\frac{\mathrm{d}v_i}{\mathrm{d}t} = 0.04v_i^2 + 5v_i + 140 - u_i + \eta + I_i, \tag{4}$$

$$\frac{\mathrm{d}u_i}{\mathrm{d}t} = a(bv_i - u_i), \tag{5}$$

where $\eta$ is the white Gaussian noise with SD = 1.0, and $I_i$ is the synaptic current. The reset of $v$ and $u$ occurs after each spike as:

$$\text{if } v_i > 30 \text{ mV, then} \begin{cases} v_i \leftarrow c \\ u_i \leftarrow u_i + d. \end{cases} \tag{6}$$

Parameters $a$, $b$, $c$, and $d$ are the internal variables of the neuron. To simplify the model and minimize the number of hyperparameters, we modeled all neurons as regular spiking neurons ($a$ = 0.02, $b$ = 0.2, $c$ = −65 mV, $d$ = 8). To solve this equation numerically, the Euler method was employed with a time resolution of 0.1 ms.

### *2.2.2. Network connectivity*

To emulate the experimental model, a hierarchically modular network with 16 modules arranged in a 4 × 4 lattice was considered. Each module contained an average of seven cells (SD = 1), and directional connections were embedded in the flow from bottom-left to top-right.

The connectivity is expressed by weight matrix $\mathbf{W} = [W]_{ij} \in \mathbb{R}^{N\times N}$, where $N$ is the total number of neurons. The probability that a connection is generated between neurons depends on whether neurons $i$ and $j$ are within the same module (intra-modular connections) or between adjacent modules (inter-modular connections). $W_{ij}$ was always set to 0, if neurons $i$ and $j$ either (1) belonged to modules that were not coupled or (2) $i = j$. For neuron pairs in the same module, connections were formed with a probability of 1. The two neurons between adjacent modules were connected with a probability of $(1 + \mathcal{D})p_{\text{inter}}$ for a coupling in the same orientation as the directional connection between modules and $(1 - \mathcal{D})p_{\text{inter}}$ for a coupling in the opposite orientation. Here, $p_{\text{inter}} \in [0, 0.5]$ is the inter-modular connection probability. The degree of directionality $\mathcal{D} \in [0,1]$ is defined for each inter-modular connections, with $\mathcal{D} = 1$ being fully compliant to the orientation of the directional connection and $\mathcal{D} = 0$ being completely random. For a given neuron pair $i$-$j$ with a synaptic connection, its weight $W_{ij}$ was chosen from a Gaussian distribution with a mean of 3 and a SD of 1. This distribution of synaptic weights was calibrated to set an excitatory postsynaptic potential of approximately 6 mV (Vogt et al., 2003).

The synaptic current injected from neuron $i$ to neuron $j$, denoted as $I_{ij}(t)$, is then given by:

$$I_{ij}(t) = \sum_{t_l<t} W_{ij} \exp\left(-\frac{t - t_l - \Delta_{ij}}{\tau}\right) H\left(t - t_l - \Delta_{ij}\right) \tag{7}$$

where $t_l$ is the time of the $l$-th spike evoked in neuron $i$, $\tau$ is the synaptic time constant set to 10

ms (Orlandi et al., 2011), and $\Delta_{ij}$ is the synaptic delay randomly assigned from a Gaussian distribution with a mean of 0.6 ms and a SD of 0.1 ms (Shimba et al., 2021). $H(\cdot)$ is a binary step function, where $H(t) = 1$ for $t > 0$ and $H(t) = 0$ otherwise. The total synaptic current injected in neuron $j$ was given by $I_j = \sum_{i=1}^{N} I_{ij}(t)$.

### 2.2.3. *Source of noise*

The noise current consists of two factors. The first factor is associated with the fluctuation in the membrane potential, described by $\eta$ in Eq. (4). The second factor, shot noise, is associated with the spontaneous release of neurotransmitters. The shot noise was a Poisson train generated with a mean of 0.01 Hz (Kavalali, 2015). When shot noise occurred in the synapse from $i$ to $j$, the current was injected into neuron $i$ at the same amplitude as that of the evoked one without synaptic delay in Eq. (7).

### 2.2.4. *Analysis of network dynamics*

To identify collective activity in the SNN model, a detection algorithm was developed based on a previous study (Orlandi et al., 2013). Starting from finding the neuron $i$ generating the first spike, we scanned for spikes from neurons receiving output connections from neuron $i$ within a time window of $2\tau$, where $\tau$ is the synaptic time constant (= 10 ms). This process was iteratively executed until no further spikes were detected, and a collective activity event was defined as an episode comprising more than 5 spikes.

To assess the synchronization among different neurons, correlation coefficients were computed from waveforms in which a spike sequence was convolved with a Gaussian kernel (SD: 25 ms). The correlation matrix for non-firing cells and cell pairs in the same module were ignored when calculating the mean correlation coefficients and functional complexity. The abovementioned setting was applied to maintain consistency with the in vitro experiments.

## 2.3. *State-transition model*

Let us consider the coarse-grained state-transition model to understand the dynamics of neurons observed in the in vitro experiment. For simplicity, this model assumes that the neurons can take only binary states: 1 (representing the activated state) and 0 (representing the resting state). Therefore, the entire system can take $2^N$ *firing states*, which are combinations of the states of $N$ neurons. Let $\vec{s}_k = (s_k(1), \ldots, s_k(N))^\top$ be a vector representing the firing states of $N$ neurons labeled by the firing state $k \in \{0, 1, \ldots, 2^N-1\}$, where $s_k(i) \in \{0, 1\}$, $(i = 0, 1, \ldots, N-1)$ and $^\top$ represents transposition of a matrix. As $s_k(i)$ are binary, one can correspond the label $k$ and the state vector $\vec{s}_k$ via the binomial expansion $k = \sum_{i=1}^{N} s_k(i) \cdot 2^{i-1}$. For example, $k = 10 = 2^{2-1} + 2^{4-1}$ in the four-node network represents the firing state $\vec{s}_{10} = (0,1,0,1)^\top$.

The network consists of $N$ nodes, each connected to its nearest neighbors, thereby simulating the modular or the hierarchically modular structure in the in vitro experiments. Let $\mathbf{W} \in [0, 1]^{N\times N}$ be the matrix which represents the connectivity between the modules. If a connection exists from node $i$ to node $j$, $W_{ij}$ and $W_{ji}$ are set to $(1+\widetilde{\mathcal{D}})/2$ and $(1-\widetilde{\mathcal{D}})/2$, respectively, where $\widetilde{\mathcal{D}} \in [0, 1]$ is the parameter defining the strength of directionality for the network.

We assume that the firing state is updated stochastically at each discrete time step, depending only on the firing state of the previous time step. We further assume that the state of node $i$ is updated depending only on $s_k(i)$ and $f_k(i)$, and is independent of other neurons. In order to describe the update of nodes in the resting state, let us introduce a vector:

$$\vec{f}_k = \mathbf{W}^{\top}\vec{s}_k = \left(f_k(1), \dots, f_k(N)\right)^{\top}, \tag{8}$$

where $f_k(i)$ denotes the synaptic input to node $i$ for a given firing state $k$. If node $i$ is in the resting state, $s_k(i) = 0$, it is activated with the probability:

$$A\left(f_k(i)\right) = \begin{cases} \sigma & \text{if } f_k(i) = 0, \\ f_k(i) & \text{if } 0 < f_k(i) < 1, \\ 1 & \text{if } 1 \le f_k(i), \end{cases} \tag{9}$$

where σ is the spontaneous activation probability and is set to $10^{-3}$ and $10^{-12}$ for the 4- and 16-node network models, respectively. The active state, $s_k(i) = 1$, persists to the following time step with probability γ and transition to the resting state with probability $1 - \gamma$, where $\gamma = 0.8$ was adopted in the present work. In summary, the transition probability of neuron $i$ in firing state $k$ is given by $P(s_{k'}(i) = 1|s_k(i) = 0) = A(f_k(i))$, $P(s_{k'}(i) = 0|s_k(i) = 0) = 1-A(f_k(i))$, $P(s_{k'}(i) = 1|s_k(i) = 1) = \gamma$, $P(s_{k'}(i) = 0|s_k(i) = 1) = 1-\gamma$. A state-transition matrix $\mathbf{T}$, whose elements $T_{k'k}$ represents the probability of transition from firing state $k$ to $k'$, is obtained by multiplying the probabilities associated with each element's state transition, as follows:

$$T_{k'k} = \prod_{i=1}^{N} P(s_{k'}(i)|s_k(i)). \tag{10}$$

A pseudo-code for obtaining $P(s_{k'}(i)|s_k(i))$ is provided in the Supplementary Material.

Once a state-transition matrix $\mathbf{T}$ is obtained, it is possible to calculate the time evolution of the probability that the system is in each firing state. Let $\vec{p}(t) = \left(p_0(t), \dots, p_{2^N-1}(t)\right)^{\top}$ be the vector, in which $k$-th component $p_k(t)$ represents the probability taking firing state $k$ at time $t$. Following the above assumptions, its time evolution is given by $\vec{p}(t) = \mathbf{T}^t\vec{p}(0)$, where $\vec{p}(0)$ is the initial condition. Specifically, the steady state $\vec{p}(\infty)$ is represented by the eigenvector of $\mathbf{T}$. To analyze the convergence of $\vec{p}(t)$, the right-hand side of the previous equation is decomposed as follows:

$$\vec{p}(t) = \sum_{i=0}^{2^N-1} c_i \lambda_i^t \vec{x}_i, \tag{11}$$

where $\lambda_i$ is an eigenvalue of **T**, $\vec{x}_i$ is the eigenvector of the corresponding eigenvalue, and $c_i$ is the decomposition coefficient. The eigenvalues and eigenvectors are ordered from the largest absolute value to the smallest. From the probability conservation property ($\sum_i T_{ij} = 1$) and the Perron-Frobenius theorem, it can be concluded that the state-transition matrices **T** always have a unique eigenvalue of 1, with all other eigenvalues with magnitudes less than 1, if **T** is irreducible. This implies that probability vector $\vec{p}(t)$ approaches $c_0\vec{x}_0$ asymptotically while other terms converge to 0 at the limit of $t \to \infty$. Since the first term ($i$ = 0) in Eq. (11), which corresponds to an eigenvalue equal to 1, represents the equilibrium probability of observing each firing state, it follows that $c_0 \sum_{i=0}^{2^N-1} x_{0,i} = 1$ holds. The eigenvalues and eigenvectors of **T** were computed using built-in MATLAB functions.

When searching for the optimal connectivity that maximizes functional complexity in a 16-node network, all topologies were categorized into specific symmetric groups to reduce the number of possible configurations. We assumed that the nodes were connected in a hierarchically modular structure with feedforward subnetworks, and that the connection between the nodes was fully directional, with $\widetilde{\mathcal{D}}$ set to 1. All the topologies were generated by rotating each of the 2×2-node groups by 0°, 90°, 180°, and 270° (see Supplementary Material Fig. S8). These assumptions reduced the sample size from $2^{20}$ ($\approx 10^6$) to $4^5$ (= 1,024). Finally, by eliminating topologically equivalent networks, the sample size was further reduced to 136. This reduction was possible because an equivalent probability vector $\vec{p}(\infty)$ that emerges from the same symmetric group result in an identical functional complexity value.

## 3. *Results*

### 3.1. *In vitro modeling with biological neurons*

#### 3.1.1. *Concept*

We conducted a series of in vitro experiments to understand how the introduction of directional connections between modules influences dynamical properties such as synchrony and integration-segregation balance in living neuronal networks. A microchannel with an asymmetrically tapered geometry (Peyrin et al., 2011; Gladkov et al., 2017) was used to implement directional connections. The asymmetric structure biases the number of axons penetrating the microchannels from the two ends because the probability of axons passing from the larger to the smaller end is higher (forward direction) than that in the opposite (reverse) direction. As shown in Fig. 1A, the overall network was designed to bear a hierarchically modular topology, wherein the network was configured in a 2 × 2 layout, with each module further subdivided into smaller 2 × 2 modules.

The entire network was designed to fit the field of view of a 20× objective lens. Asymmetric channels were used to couple the small modules and arranged such that the network had a pseudo-feedforward flow of activity from bottom-left to top-right. This structure was adopted based on the hypothesis that since excitatory neurons are the dominant population in cultured cortical neurons (Kono et al., 2016), breaking recurrent connections would reduce excessive synchronization and enrich spatiotemporal dynamics in neuronal networks. By comparing the network with symmetric microchannels, we investigated the effects of directional connections.

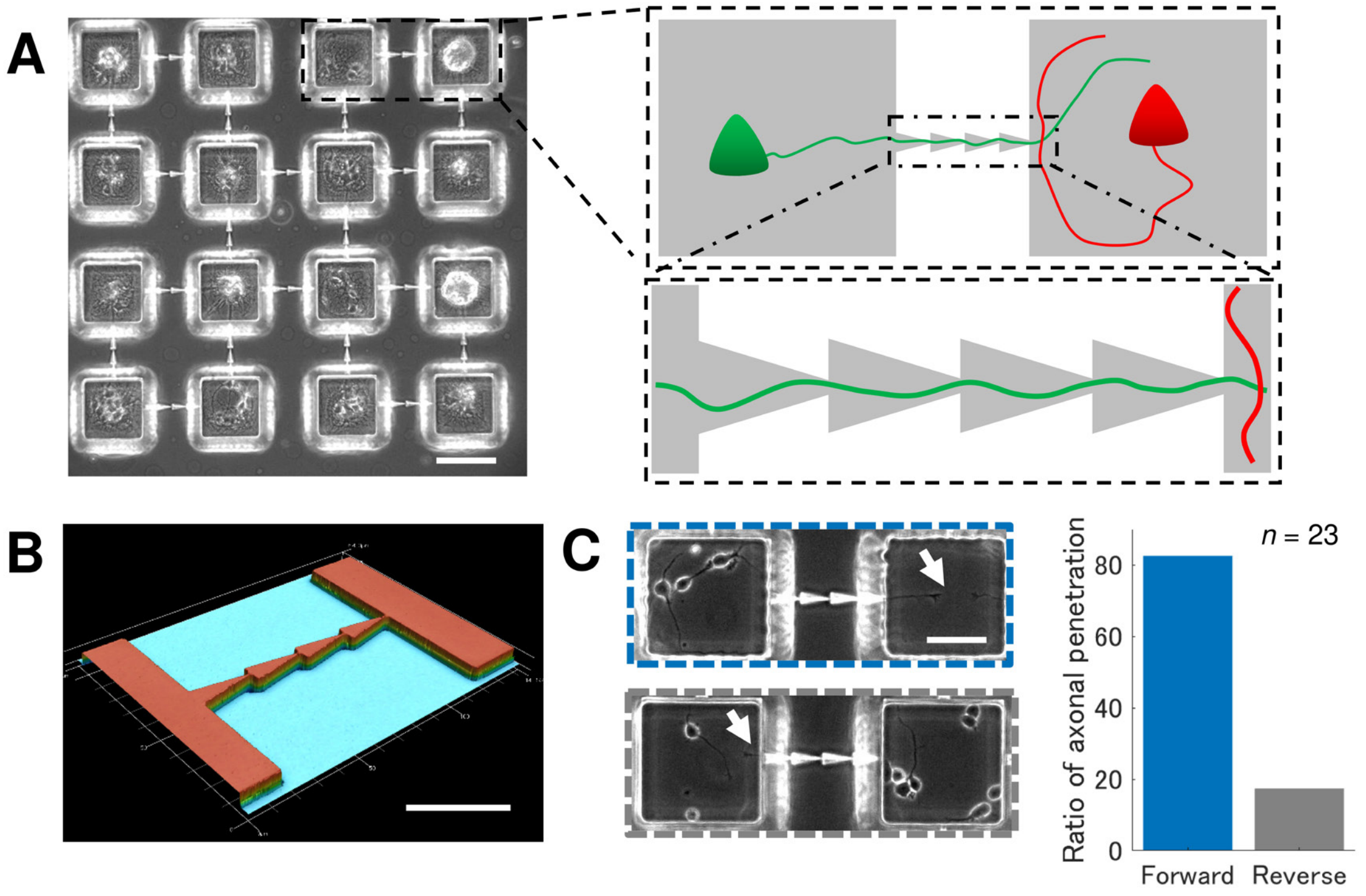

Fig. 1. Reconstruction of a modular neuronal network with directional connections. (A) Left: Phase contrast image of the cultured neuronal network grown in a microfluidic device with directional intermodular connections. This network has a pseudo-feedforward flow from bottom-left to top-right. Right: Schematic illustration of oriented axonal elongation in asymmetric microchannels. (B) Confocal micrograph of the master mold used to fabricate asymmetric microchannels. (C) The elongation of neurite in the forward direction (from long side to short side) (top) and in the reverse direction (bottom). Bar graph represents the ratio of axons penetrating in forward and reverse direction ($n = 23$). Scale bars, 100 μm (A) and 50 μm (B, C).

*3.1.2. Controlling directionality with asymmetric microchannels*

Fig. 1B shows a confocal micrograph of the asymmetric microchannel in the master mold. The widths of the larger and smaller ends were measured to be 9.8 ± 0.3 and 2.4 ± 0.5 µm, respectively (mean ± SD, $n$ = 5). The symmetric channels had an average width of 6.1 ± 0.5 µm ($n$ = 5). For both types of microchannels, the average height was 1.4 ± 0.0 µm ($n$ = 10).

Next, we examined whether the asymmetric microchannels biased the probability of axons penetrating in the intended direction. Evaluation of the probability using time-lapse imaging revealed that while axons penetrated in both the forward and reverse directions (Fig. 1C), the percentage of axons penetrating in the forward direction was 82.6%. These results indicate that asymmetric microchannels bias the probability of penetrating neurites and realize directional connectivity.

We further investigated whether the direction of neural activity propagation was biased by the asymmetric channel, using fluorescence calcium imaging (Figs. 2A, 2B). This analysis revealed that the probability of propagation in the reverse direction was significantly lower than that in the forward direction (Fig. 2C). Such a bias was only observed in the asymmetric microchannel (mean ± SEM, forward, 91.0 ± 0.8%; reverse, 59.8 ± 1.6%; $n$ = 24) and not in the symmetric channel (forward, 84.3 ± 1.0%; reverse, 87.6 ± 0.7%; $n$ = 20). Moreover, asymmetric channel induced a greater propagation delay in the reverse direction (forward, 54 ± 1 ms; reverse, 115 ± 5 ms), an effect that was not present in the symmetric channel (forward, 62 ± 2 ms; reverse, 85 ± 3 ms) (Fig. 2C). The asymmetric microchannels thus induced functional directionality by suppressing the propagation probability and increasing the propagation delay in the reverse orientation.

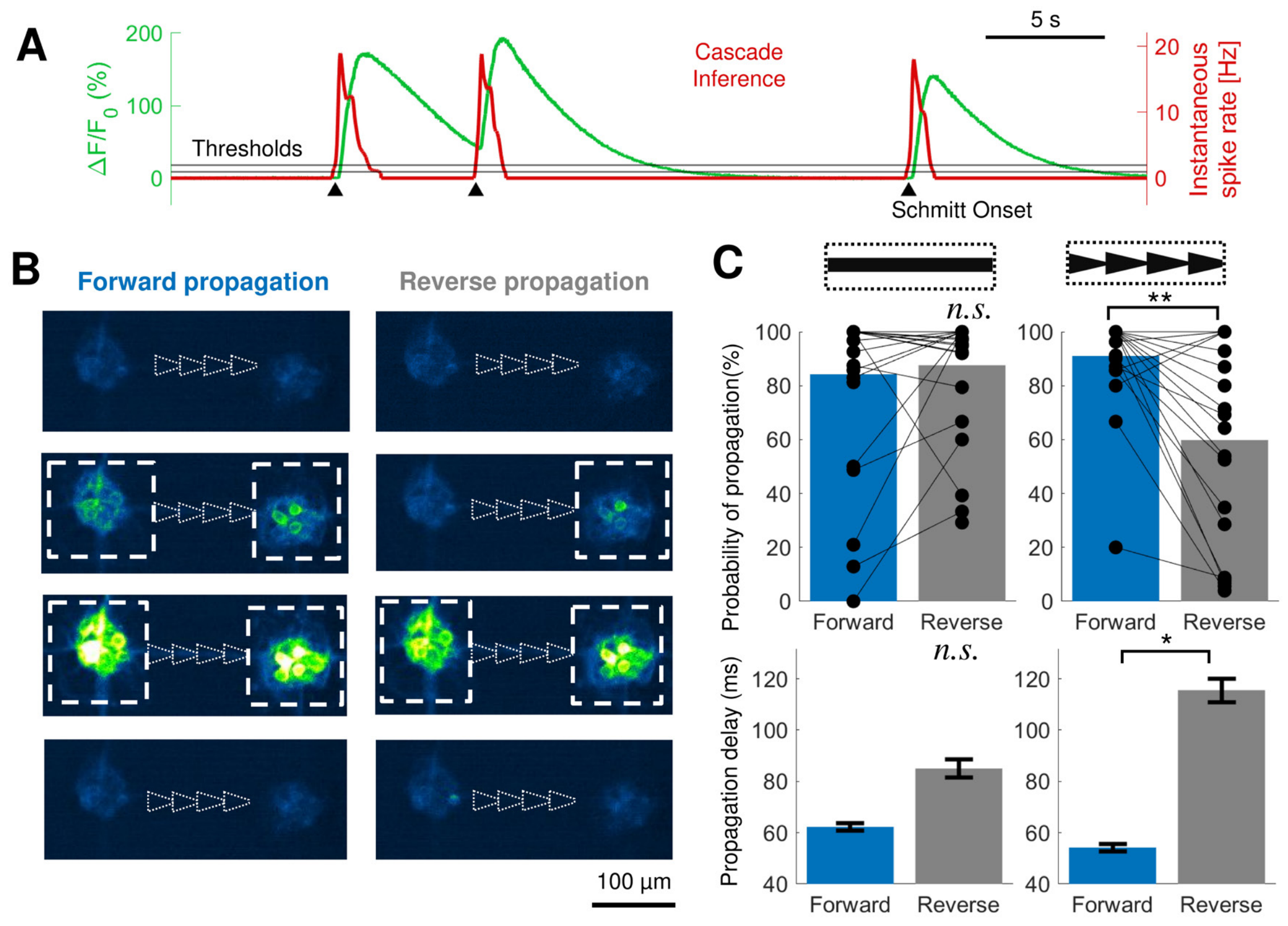

Fig. 2. Signal propagation characteristics of asymmetric microchannels. (A) The spike inference from fluorescence calcium imaging waveforms. Vertical axes are the relative fluorescent unit $\Delta F/F_0$ (green) and the inferred spike rate (red). The onset of firing was detected by Schmitt-trigger method. (B) The timelapse of activity propagation. Left: Forward propagation. Right: Reverse propagation. While the activity propagates quickly in the forward direction, the activity propagates in the reverse direction with a larger delay. (C) Analysis of the property of signal propagation in symmetric and asymmetric channels ($n$ = 20, 24). Top: Probability of activity propagation. Bottom: propagation delay. Blue and gray bar represents the mean value in forward and reverse direction, respectively. mean ± standard error of the mean (SEM) $*p < 0.05$; $**p < 0.01$; n.s., not significant (two-tailed Student's $t$-test).

### *3.1.3. Impact of directional connections on spontaneous activity patterns*

The spontaneous activity of cultured neuronal networks is characterized by a highly synchronized bursting activity, often referred to as network bursts (Orlandi et al., 2013). To investigate how micropatterning the neurons in a modular geometry and introducing asymmetry in intermodular connections affect the spatiotemporal dynamics of the neuronal network, we constructed two types of networks and compared their spontaneous activity patterns at 10 DIV (Fig. 3A). As shown in previous studies (Yamamoto et al., 2018a; Takemuro et al., 2020; Yamamoto et al., 2023), patterning primary neurons in a modular geometry enabled the network to exhibit locally synchronized activities that coexisted with globally synchronized ones (Fig. 3A1; non-directional network). When the modules were connected with asymmetric channels, globally synchronized activity was further suppressed, and local activity occurred more frequently with greater combinations of modules (Fig. 3A2; directional network). Furthermore, a comparison of globally synchronized activities between the two networks revealed that the latency for the activity to propagate through the entire network was greater in the network with asymmetric channels (Fig. 3A2, inset). This suggests a decrease in synchrony, even over a timescale as large as 10 ms.

To quantify the changes caused by the introduction of asymmetric channels, we compared the GNA, mean correlation coefficient, and functional complexity of the two networks. The above-mentioned decrease in globally synchronized activity could be quantified based on the distribution of GNA, whose median was reduced by 34.4% in the non-directional network. This in turn spread the value of the correlation coefficients between 0 (no correlation) and 1 (globally correlated) (see the bottom left panels in Figs. 3A1 and 3A2). Consequently, the mean correlation coefficient was reduced by 17.7% and the functional complexity increased by 31.7% in the directional networks (Fig. 3B). The results show that embedding asymmetric microchannels is effective in suppressing excessive synchronization and balancing the integrated and segregated states.

Finally, to better understand the mechanisms underlying the suppression of synchrony, the propagation of activity between modules was evaluated using cross-correlation functions between neurons. The cross-correlation functions and propagation delays accumulated across all samples are summarized in Fig. 3C. For neuron pairs affiliated with neighboring modules ($h = 1$), a single asymmetric channel facilitates propagation in the forward direction and causes a larger delay when the activity propagates in the reverse direction (Fig. 2). Consequently, the cumulative distribution function was greater in the directional network for negative delays (corresponding to reverse propagation) and in the non-directional network for positive delays (corresponding to forward propagation) (Fig. 4C, left). For neuron pairs between the bottom-left and top-right modules ($h = 6$), the cumulative distribution function of the directional network was biased towards positive delays compared to the non-directional network (Fig. 4C, left) in both the positive and negative delays. This implies that the fraction of reverse propagation is further reduced when the activity propagates through a larger number of asymmetric microchannels. The cross-correlation analysis for all hops is provided in the Supplementary Material (Fig. S3).

In summary, we fabricated modular networks with asymmetric microchannels and found that asymmetric microchannels inhibit the probability and increase the delay of reverse propagation, consequently suppressing excessive synchrony in cultured neuronal networks.

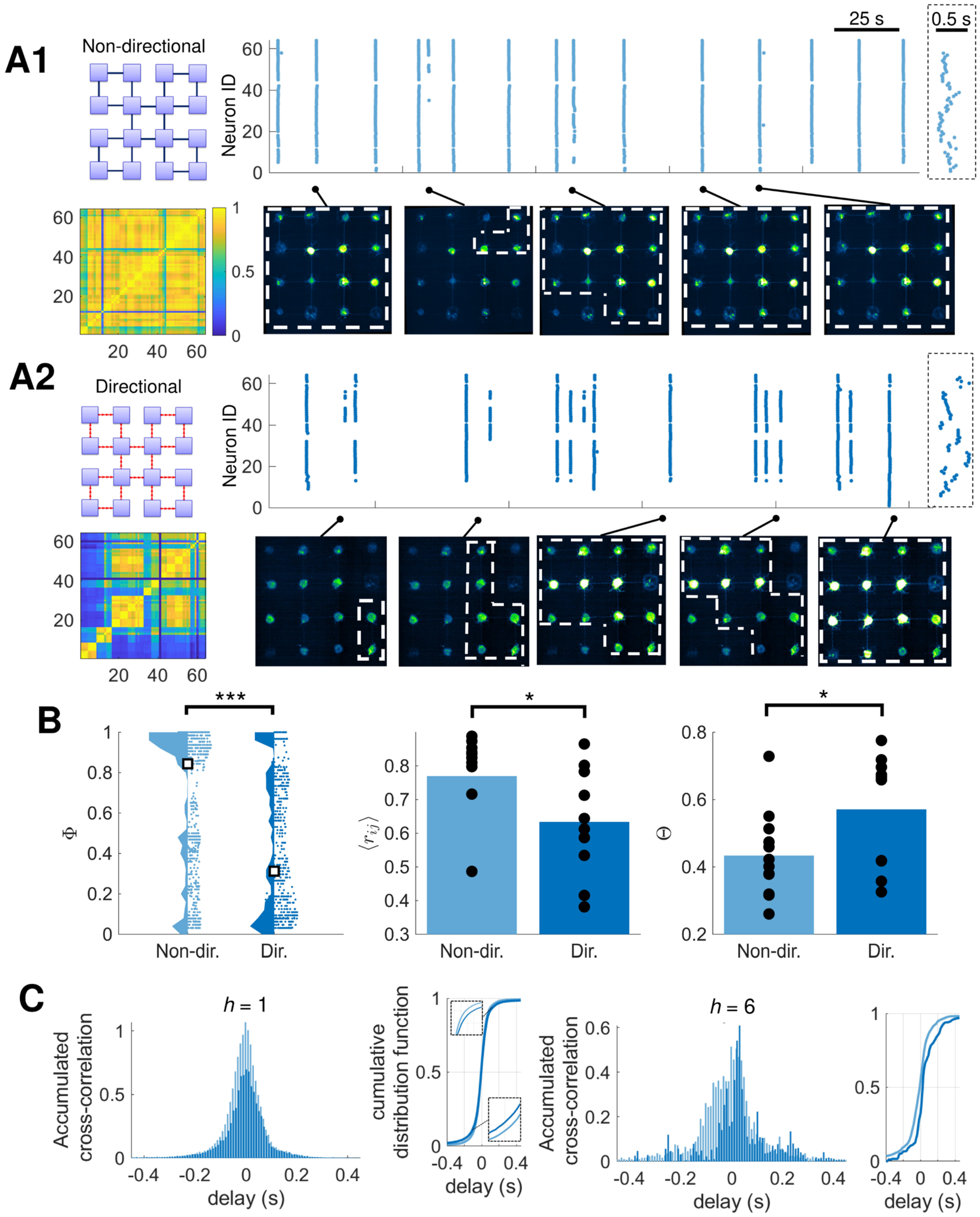

Fig. 3. Spontaneous activity recorded by fluorescent calcium imaging and quantification of its dynamics. (A) Representative examples of spontaneous activity in Non-dir. (non-directional) and Dir. (directional) networks. Left: A schematic illustration of the overall network structure and the correlation matrix obtained from recorded activity. Right: Raster plot of the spontaneous activity and snapshots of GCaMP6s fluorescence at representative time points. (B) Statistics of the spontaneous activity in non-directional and directional networks ($n$ = 12, 10). Left: Global network activation (GNA): Φ. Violin plots (left side) represent the probability distribution of GNA

observed across all networks, and individual value are shown in swarm charts (right side). Black boxes indicate the medians. Middle: Mean correlation coefficient. Right: Functional complexity. *$p < 0.05$ (two-tailed Student's *t*-test) (C) Analysis of propagation delay using cross-correlation function $\langle \tilde{R}_{ij}(\tau) \rangle_h$. Left: $h = 1$, right: $h = 6$.

### *3.2. In silico modeling with spiking neurons*

#### *3.2.1. Concept*

To explore the underlying mechanisms responsible for the enhanced dynamical complexity resulting from directional intermodular connections, we built a spiking neural networks (SNN) model that emulated in vitro experiments. Fig. 4A shows the SNN model that replicates the hierarchically modular neuronal network with asymmetric channels (see Supplementary Material for details). In the SNN model, the degree of asymmetry in intermodular coupling, analogous to the biasing ratio of axonal penetration from the two ends, was parameterized by directionality $\mathcal{D}$. The densities of intermodular couplings, corresponding to the total number of axons penetrating through a microchannel, were controlled via intermodular connection probability $p_{\mathrm{inter}}$. The probability of intramodular connection was fixed as 1. Spontaneous activity was triggered by Poisson noise within each neuron. The effect of directional intermodular connections on network dynamics were characterized by evaluating GNA, mean correlation coefficient, and functional complexity from the simulated activity patterns.

#### *3.2.2. Effect of directional intermodular connections on network dynamics*

We first simulated spontaneous activity at two extreme degrees of directionality ($\mathcal{D} = \{0, 1\}$), with $p_{\mathrm{inter}}$ set to 0.1 (approximately eight axons between two modules), and examined the difference in network dynamics (Fig. 4B). In the case of $\mathcal{D} = 0$, corresponding to non-directional networks, both globally and locally synchronized bursting events coexisted in a single network. Conversely, when $\mathcal{D} = 1$, globally synchronized activity was completely suppressed, and various patterns of locally synchronized bursting was observed. The duration of bursting events was also found to decrease compared to the non-directional ($\mathcal{D} = 0$) networks, due to the decrease in the fraction of neurons involved in each event (Supplementary Material Fig. S5). The observed trend of reduced burst duration aligns with the findings from biological experiments (Fig. 3A insets, Supplementary Material Fig. S1).

To understand the mechanisms by which directional connections increase the variability of the synchronous bursting events, we quantified GNA, neural correlation, and functional complexity for non-directional ($\mathcal{D} = 0$) and fully-directional ($\mathcal{D} = 1$) networks. As summarized in Fig. 4C, fully-directional connections suppress the synchronization and balance the integration-segregation state. GNA peaked at around 1 and 0.02 for networks with $\mathcal{D} = 0$ but did not attain a value of 1 for networks with $\mathcal{D} = 1$. This is because reverse propagation between adjacent modules was blocked by the directional connections. This change in the distribution of GNA, caused by making the intermodular connections directional, decreased the mean correlation coefficient by 46.7% ($n$ = 40 network realizations, $p = 6.34 \times 10^{-15}$) and increased the value of functional complexity by 74.2% ($n = 40$, $p = 1.00 \times 10^{-19}$).

As a general trend, increasing the directional bias of the intermodular connections through the parameter $\mathcal{D}$ decreased neural correlation and increased the value of functional complexity (Fig. 4D). For a network with $p_{\text{inter}} = 0.1$, the mean correlation significantly decreased when $\mathcal{D}$ exceeded 0.6. The value of functional complexity was more sensitive to the variation in $\mathcal{D}$ and exhibited a significant increase when $\mathcal{D}$ exceeded 0.2.

The specific value of $\mathcal{D}$ at which the mean correlation and functional complexity diverge from the network characterized by $\mathcal{D} = 0$ was dependent on the intermodular connection probability $p_{\text{inter}}$. When $p_{\text{inter}}$ was too high to render the network non-modular, both mean correlation and functional complexity were insensitive to the change in $\mathcal{D}$, until $\mathcal{D}$ approached 1. When $p_{\text{inter}}$ was set to a low value (= 0.05) and the number of intermodular connections were too sparse, both mean correlation and functional complexity remained insensitive to changes in $\mathcal{D}$. The mean correlation and functional complexity obtained for other values of $p_{\text{inter}}$ and $\mathcal{D}$ are summarized in Fig. S4. In summary, directionality and modularity interact jointly to enrich the variability of spontaneous activity and balance integration-segregation in neuronal networks.

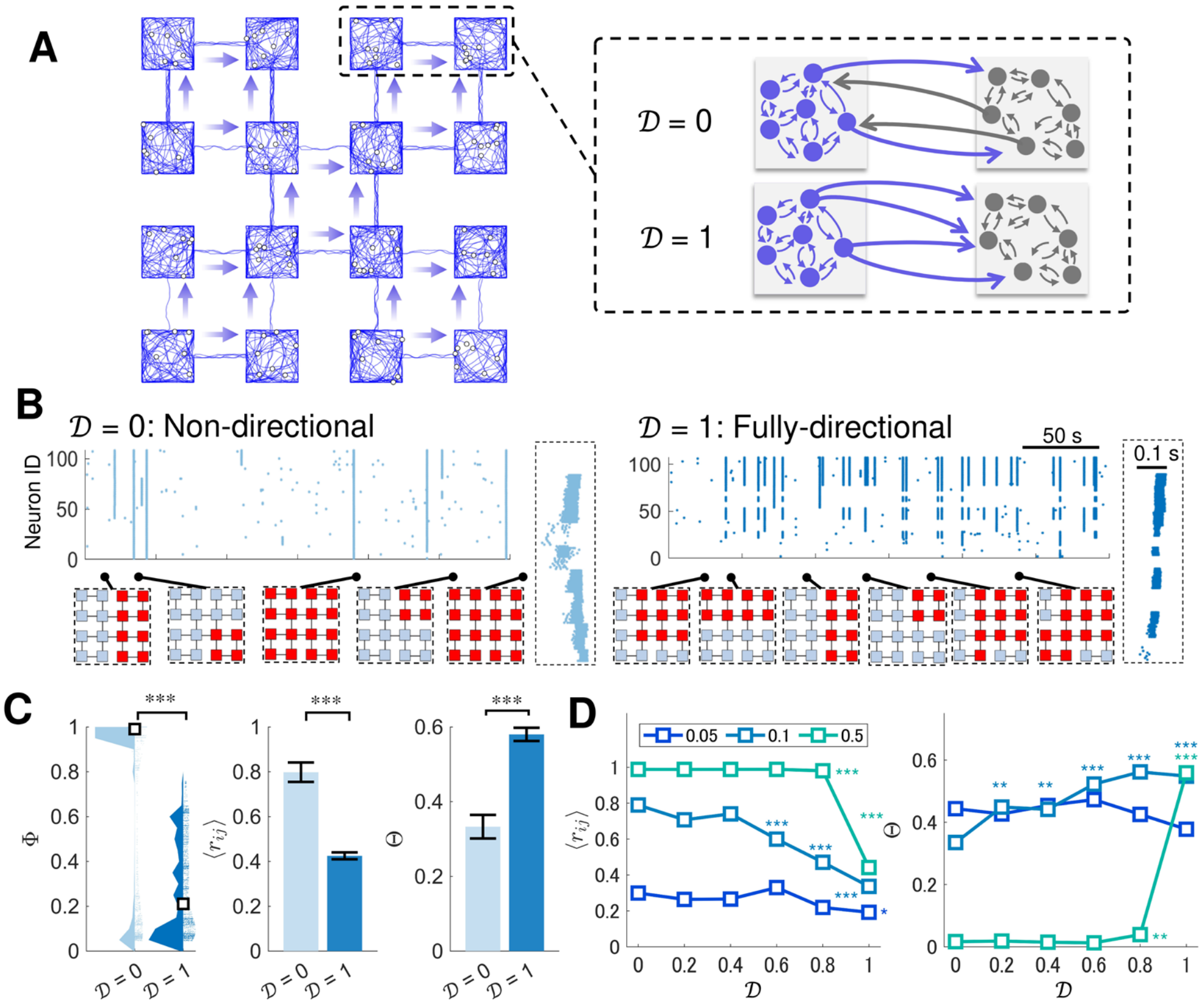

Fig. 4. Numerical simulation of spontaneous activity using SNN model. (A) Illustration of the network model with hierarchically connected 4 × 4 modules. Each module contains a mean of 7 neurons. There is a pseudo-feedforward flow from the bottom-left module to top-right module. Connection formed between modules is controlled by the directionality $\mathcal{D}$. (B) Raster plots and corresponding spatial firing patterns in non-directional and fully-directional networks. The insets show a close-up of representative burst events with large GNA. (C) Statistics of the spontaneous activity in non-directional and fully-directional networks. Left: Distribution of GNA ($\Phi$), representing median value with the square dot. Middle: Mean correlation coefficient $\langle r_{ij} \rangle$. Right: Functional complexity $\Theta$. ($n$ = 40, mean ± SEM. *$p < 0.05$; **$p < 0.01$; ***$p < 0.001$; n.s., not significant, two-tailed Student's $t$-test) (D) The dependence of mean correlation and functional complexity on $\mathcal{D}$ and $p_{\text{inter}}$. Asterisks represent statistical difference against the value at $\mathcal{D} = 0$ (*$p < 0.05$; **$p < 0.01$; ***$p < 0.001$).

*3.3. Analytical formulation with state-transition model*

*3.3.1. Concept*

We here formulate a simple state-transition model to show that the relationship between the intermodular connectivity and the degree of network synchrony can be derived analytically. In this model, the transition of network firing state from $\vec{s}_k$ to $\vec{s}_{k'}$ is given by the transition probability $T_{k\prime k}$. For a given network defined by a matrix $\mathbf{T} = [T]_{k'k} \in \mathbb{R}^{2^N \times 2^N}$, where $N$ is the number of nodes, properties of its equilibrium dynamics are predicted by calculating the eigenvectors and eigenvalues of **T** (see Section 2.3). Here, neurons within the same module are represented as a single dynamical node transitioning between resting and activated states. This abstraction is validated by the observation in biological experiments and SNN models that neurons affiliated to the same module were mostly activated simultaneously (Figs. 3A, 4B). As shown in Fig. 5A, the resting node can be activated through synaptic input $f_k(i)$ or spontaneously with a probability σ. To ensure the irreducibility of **T**, we assume that the active state persists to the following time step with a probability γ, independent of synaptic input $f_k(i)$. Further details are provided in Section 2.3.

Various parameters explaining network dynamics can be estimated from the eigendecomposition of **T**. For example, the expected number of active nodes, which corresponds to the value of the GNA calculated in the previous two sections, is given by

$$\langle \widetilde{\Phi} \rangle = \sum_{k=1}^{2^N-1} \left( \frac{p_k(\infty)}{1 - p_0(\infty)} \right) \widetilde{\Phi}_k , \tag{14}$$

where $p_k(\infty)$ is the equilibrium probability of observing the firing state $\vec{s}_k$, and $\widetilde{\Phi}_\mathrm{k} = \sum_{i=1}^{N} s_k(i)$ is the number of nodes that is active in a given firing state $\vec{s}_k$. The firing state index $k = 0$ was omitted from the summation because it corresponds to a firing state with no activity. Importantly, $\vec{p}(\infty) = \left(p_0(\infty), \cdots, p_{2^N-1}(\infty)\right)^\mathrm{T}$ is derived analytically from **T** as its normalized eigenvector corresponding to an eigenvalue of 1, as is shown below: The probability vector of observing each firing state at time $t$ is related to the state-transition matrix **T** as $\vec{p}(t+1) = \mathbf{T}\vec{p}(t)$ . At $t \to \infty$, $\vec{p}(t+1)$ always converges to $\vec{p}(\infty)$, the equilibrium probability of the network (see Section 2.3). Because $\mathbf{T}\vec{p}(\infty)$ equals $\vec{p}(\infty)$ at equilibrium, $\vec{p}(\infty)$ is the eigenvector of the state-transition matrix **T** for an eigenvalue of 1.

The correlation coefficient between nodes $i$ and $j$, denoted as $\tilde{r}_{ij}$, is given by

$$\tilde{r}_{ij} = \frac{\sigma_{ij}}{\sqrt{\sigma_{ii}\sigma_{jj}}}, \tag{15}$$

where $\sigma_{ij} = \sum_{k=0}^{2^N-1} p_k(\infty)\big(s_k(i) - \bar{s}(i)\big)\big(s_k(j) - \bar{s}(j)\big)$ is the covariance between nodes $i$ and $j$, $\bar{s}(i) = \sum_{k=0}^{2^N-1} p_k(\infty) s_k(i)$ is the expected degree of the activation of node $i$, and $s_k(i)$ is the state of node $i$ in the $k$-th network firing state. Starting from $\tilde{r}_{ij}$, the mean correlation $\langle \tilde{r}_{ij} \rangle$ and functional complexity $\widetilde{\Theta}$ can be calculated similarly using the equations used in biological experiments (see Section 2.1.7).

### *3.3.2. Four-node network model*

To better illustrate this model, we derived $\langle \widetilde{\Phi} \rangle$, $\langle \tilde{r}_{ij} \rangle$, and $\widetilde{\Theta}$ for two simple networks possessing four nodes: one without any connection (Supplementary Material Fig. S6(a)) and one with connections in a feedforward manner (Fig. 5B), referred to as *isolated* and *feedforward* networks, respectively. The state-transition matrices **T** for the two networks are shown in Supplementary Material Fig. S6(b).

In the *isolated* network, there were only five firing states that exhibited a non-negligible equilibrium occurrence probability ($p_k(\infty) > 10^{-4}$). The five firing states were either firing states with one active node ($\vec{s}_1$, $\vec{s}_2$, $\vec{s}_4$, $\vec{s}_8$) or a firing state with no active nodes ($\vec{s}_0$) (Fig. S6(a)). In contrast, in the *feedforward* network, $p_k(\infty)$ was more widely distributed across multiple firing states. Specifically, the occurrence probability exceeded $10^{-4}$ for all firing states, except $\vec{s}_3$, $\vec{s}_5$, $\vec{s}_9$, which corresponded to firing states with two active upstream nodes (1 and 2, or 1 and 3) or two diagonal nodes (1 and 4) (Fig. 5B). These results can be understood intuitively by considering that nodes cannot be activated synchronously without connections, whereas they can be activated both synchronously and asynchronously when they are connected. Furthermore, in the *feedforward* network, the occurrence probabilities of $\vec{s}_3$, $\vec{s}_5$, and $\vec{s}_9$ were low because the activation of node 1 was likely to trigger the activation of nodes 2 and 3. We note that the absolute values of the probability distribution depend on **T**, as well as σ and γ.

The correlation coefficients $\tilde{r}_{ij}$ calculated for the two networks are shown in Figs. 5B and S6(a). For the *isolated* network, the expected number of firing nodes in a given event $\langle\tilde{\Phi}\rangle$ was calculated to be 1.01, indicating that individual nodes activate mostly independently. Correspondingly, all non-diagonal components of the correlation matrix were approximately 0 ($<10^{-16}$), resulting in a functional complexity of 0. Conversely, for the *feedforward* network, expected number of firing nodes $\langle\tilde{\Phi}\rangle$ was 1.61, indicating an increased fraction of co-activations. The non-diagonal components of correlation matrix were spread in the range of 0.2 and 0.45 with a mean of $\langle\tilde{r}_{ij}\rangle$ = 0.39, leading to a functional complexity of 0.11. In short, the state-transition model analytically formulated network dynamics depend on its structure by employing various parameters that correspond to the measurements used in cultured neuronal networks and SNN models.

### *3.3.3. Sixteen-node network model*

To analytically derive how directional connections disrupt global synchrony and increase the dynamical complexity in modular networks with directional connections, we constructed a state-transition model with 16-nodes (Fig. 5C, left). Here, the nodes were connected in a hierarchical and pseudo-feedforward manner, emulating the network structure discussed in both the in vitro experiment and the SNN model. The directionality between nodes was parameterized with $\tilde{\mathcal{D}} \in [0,1]$, where $\tilde{\mathcal{D}} = 0$ and $\tilde{\mathcal{D}} = 1$ represents non-directional and fully-directional couplings, respectively. The spontaneous activation probability was set to $\sigma = 10^{-12}$, ensuring the mean correlation of the network with $\tilde{\mathcal{D}} = 0$ approaches to that of the SNN model with $\mathcal{D} = 0$ ($\langle\tilde{r}_{ij}\rangle$ =

0.802 and $\langle r_{ij} \rangle$ = 0.798). Correlation matrices for $\widetilde{\mathcal{D}}$ = {0, 0.2, 0.4, 0.6, 0.8, 1} are depicted in Fig. S7.

As summarized in Fig. 5C, the dependences of $\langle \widetilde{\Phi} \rangle$, $\langle \tilde{r}_{ij} \rangle$, and $\widetilde{\Theta}$ on directionality $\widetilde{\mathcal{D}}$ were consistent with the results of SNN models. As $\widetilde{\mathcal{D}}$ increases, the expected number of active nodes $\langle \widetilde{\Phi} \rangle$ decreased, which in turn reduced the correlation coefficient $\tilde{r}_{ij}$. Specifically, increasing $\widetilde{\mathcal{D}}$ from 0 to 1 reduced the values of $\langle \tilde{r}_{ij} \rangle$ and $\langle \widetilde{\Phi} \rangle$ by 71.5% and 74.6%, respectively, and increased $\widetilde{\Theta}$ by 900.0%. Hence, the state-transition model accurately predicts network behaviors, including synchronization and complexity, without the need for a detailed numerical simulation of the spiking activity, as in the SNN model.

*3.3.4. Optimal configuration search*

Evaluation of the functional complexity of the 16-node network required ~45 min on a desktop PC equipped with an Intel i7-12700 CPU and an NVIDIA GeForce RTX 3060 GPU. The duration is 8.9 times faster than the time required by the SNN simulation to compute a 20 min-long session of spontaneous activity in 40 networks of 112 neurons. Taking advantage of this property of the state-transition model, we sampled network topologies with various combinations of directional connections and searched for an optimal configuration predicted by the state-transition model. Considering the symmetries of the connection topologies (see Section 2.3 for details), we reduced the sample size to 136 and assessed the functional complexity $\widetilde{\Theta}$ across all configurations.

The optimal configuration with the highest $\widetilde{\Theta}$ is illustrated in Fig. 5D. The optimal topology closely resembled the original topology (Fig. 5C) and was obtained by rotating nodes 9–12 by 270°. The correlation coefficients $\tilde{r}_{ij}$ were widely distributed, ranging from approximately 0 to 0.7, which led to a 21.7% increase in functional complexity compared to the original topology (Fig. 5E). Conversely, a significant rewiring from the original topology was required to obtain the topology with the lowest $\widetilde{\Theta}$, which is also illustrated in Fig. 5D. The $\tilde{r}_{ij}$ in this network peaked bimodally at approximately 0 and 0.4, leading to a decrease in functional complexity. Compared to the original topology, the functional complexity $\widetilde{\Theta}$ was 51.7% lower in this topology (Fig. 5E).

To validate the results of state-transition model, the spontaneous activity in the abovementioned topologies was simulated using the SNN model and the functional complexity Θ was computed from the spontaneous activity patterns. As summarized in Fig. 5F, the trend obtained in the SNN model agreed well with the prediction of the state-transition model. In the topology with the lowest $\widetilde{\Theta}$, the functional complexity Θ decreased significantly by 38.9%. Contrarily, comparing the topology with the highest $\widetilde{\Theta}$ against that of the original one, the mean value of Θ was higher by 4.0%, but the difference was statistically insignificant. The discrepancy between the SNN and the state-transition model could be due to the nonlinearity of the nodes, which is considered in the former but not in the latter. The results show that the state-transition model serves as a zeroth approximation for predicting the functional complexity in a modular network with a given connectivity pattern and validates the use of the directional topology explored in the wet experiments (Fig. 3A2).

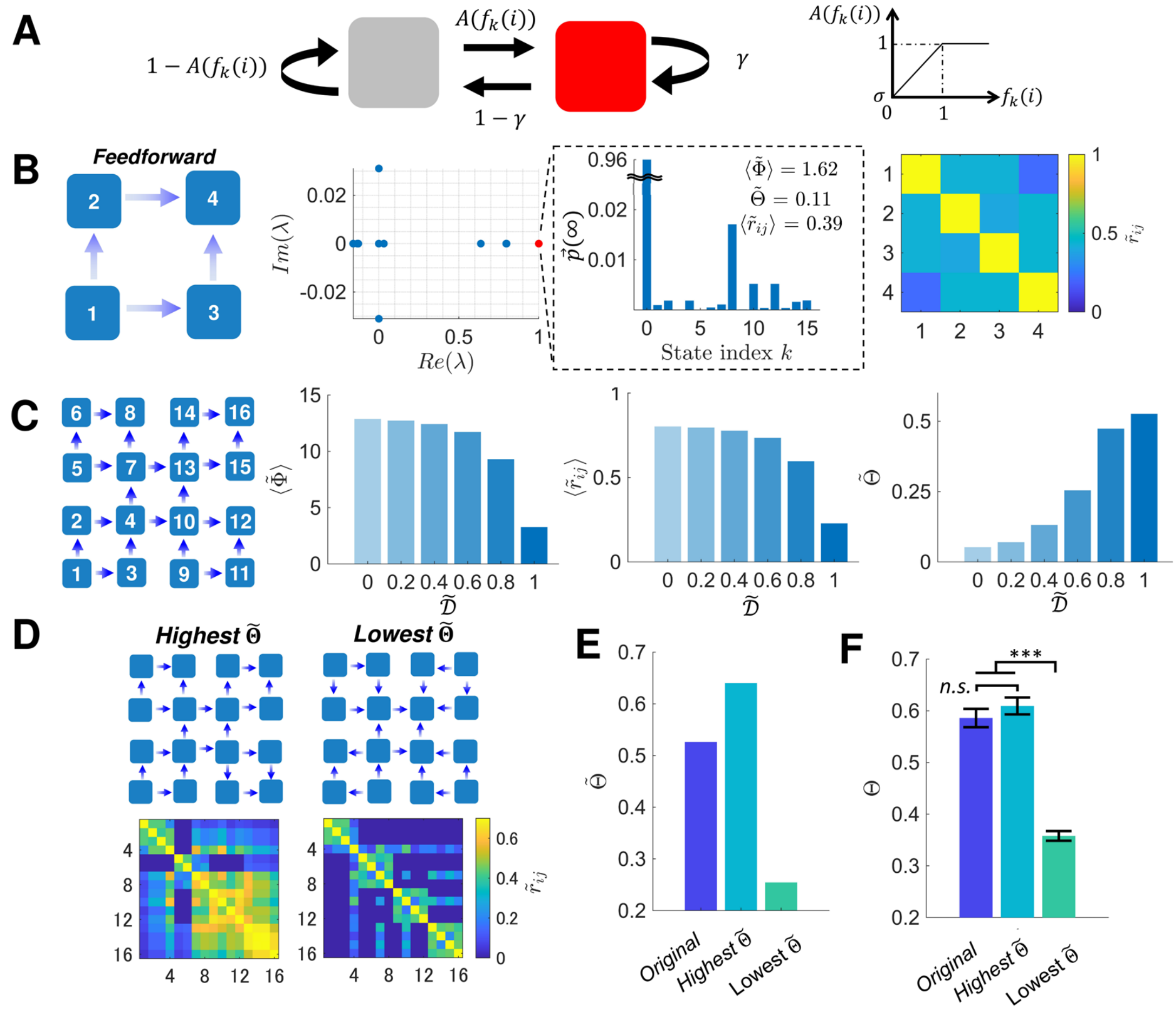

Fig 5. The probabilistic state-transition model. (A) Transition between the inactive (gray) and active (red) states depending on the synaptic input $f_k(i)$, spontaneous activation $\sigma$, and persistence probability $\gamma$. (B) Representative analysis of a 4-node network. Left: The connection structure of the *feedforward* network. Middle: Eigenspectrum obtained from the state-transition matrix **T** and the equilibrium probability of observing network firing states $\vec{p}(\infty)$, derived from the eigenvector corresponding to an eigenvalue of 1. Statistics such as the expected number of active nodes $\langle\tilde{\Phi}\rangle$, mean correlation $\langle\tilde{r}_{ij}\rangle$, and functional complexity $\tilde{\Theta}$ are also displayed. Right: The correlation matrix $\tilde{r}_{ij}$. (C) The dependence of $\langle\tilde{\Phi}\rangle$, $\langle\tilde{r}_{ij}\rangle$, and $\tilde{\Theta}$ on directionality $\tilde{\mathcal{D}}$ in a 16-node network. (D) Topologies which maximize and minimize the value of $\tilde{\Theta}$. Top: Illustration of the topologies. Bottom: Correlation matrices. (E) The value of functional complexity in the original, highest $\tilde{\Theta}$, and lowest $\tilde{\Theta}$ topologies. (F) Simulation of spontaneous activity using SNN model to verify the prediction from the state-transition model. $p_{\mathrm{inter}}$ and $\mathcal{D}$ were set to 0.1 and 1, respectively. $n = 40$ $^{***}p < 0.001$; (two-tailed Student's $t$-test).

## *4. Discussion*

In this study, we used cultured neuronal networks grown in microfluidic devices, along with SNN and state-transition models to investigate how directional connections in modular neuronal networks affect network dynamics. In cultured neuronal networks, implementing pseudo-feedforward connectivity effectively suppressed excessive synchronized activity and increased the integration-segregation balance in spontaneous activity patterns. The SNN models showed that directionality and modularity cooperatively enhanced the integration-segregation balance. Finally, we showed that the structure-function relationships in neuronal networks can be exploited using the state-transition model, which enables an analytical search for an optimal connection structure.

Our results imply that directional connections in a neuronal network play additional roles beyond the generation of information flow. Previously, Izhikevich showed that the precise tuning of directional connections and propagation delays enriched spatiotemporal firing patterns, even in a small network, by employing the SNN model (Izhikevich, 2006). Using abstract neuron models, Curto et al. demonstrated that the number of attractors in a small network can be diversified by precisely tuning the flow of directional connections (Curto & Morrison, 2019). We demonstrated how this concept extends to modular neuronal networks and validated the theoretical predictions in corresponding experiments using cultured neurons.

In wet experiments using cultured neurons, directional connections between individual modules were realized using asymmetric microchannels that biased the probability of neurite penetration at both ends. Reconstitution of directional connections, sometimes referred to as "axon diodes," is an active field of study. In previous studies, various designs of microchannels were proposed such as funnel (Peyrin et al., 2011), barbed (le Feber et al., 2015), arche (Renault et al., 2016), multi-taper (Gladkov et al., 2017), and stomach (Forró et al., 2018) geometries. Additionally, other methodologies such as sequential seeding in two chambers (Pan et al., 2015) and self-organization of neuronal populations (Brewer et al., 2013), have been used to control directionality. Multi-tapered microchannels have been explored previously by Gladkov et al. in microchannels of approximately 500 μm. In the present work, we have downsized the microchannel lengths to 100 μm in order to reduce the network area and ensure scalability.

The asymmetrical microchannels not only biased the directionality of the signal propagation but also increased propagation delay in the reverse orientation (Fig. 3A). This observation agrees with previous studies that used barbed and funnel channels (le Feber et al., 2015; Renault et al., 2015). This effect is attributed to fewer axons extending in the reverse direction, consequently increasing the time required to transmit a sufficient number of spikes to activate the downstream module.

The state-transition model proposed in our study provides a novel framework to theoretically investigate the structure-function relationships in a complex network of nonlinear excitable units that behave in a stochastic manner. Neuron dynamics can be described by deterministic equations; thus, analytical methods developed for deterministic dynamic systems have been frequently applied to analyze neural dynamics. In particular, structure-function relationships in deterministic models have been explored in complex networks of phase oscillators (Ulloa Severino et al., 2016; Yamamoto et al., 2018b). Although deterministic models can describe the detailed dynamics of neurons, e.g., spikes, it is difficult to analyze the interplay between network topology and the functionality of the system, owing to its nonlinearity. For example, a method of master stability function (Pecora & Carroll, 1998 and Fujisaka & Yamada, 1983) is often employed for this purpose because it provides the relations between the eigenvalues of the Laplacian matrices of networks and the Lyapunov exponent of the synchronized oscillations. However, this method is only applicable to complete synchronized states, which are not useful for information processing. In contrast, the state-transition model approach simplifies the dynamics of neurons, making it easier to represent the influence of the network structure on the dynamics. Note that such a state transition between discretized states can be modeled using deterministic models (Garcia et al., 2012). Our model considers the presence of stochasticity, which is inherent in real systems, and enables us to consider transitions that do not frequently occur and are neglected by deterministic models. Indeed, in this study, we have shown that this model predicts the mean correlation and functional complexity, which roughly agreed with in vitro experiments and SNN models.

Although the present study provides a novel platform for integrating in vitro experiments and computational/mathematical models, it has some limitations. In the wet experiment, the asymmetric microchannel could not reconstitute a fully directional connection, and the effect of the directional connections diminished as the microchannel matured (Fig. S2). Such an issue has also been pointed out by other research groups attempting to realize the microfluidic control of axon directionality (Renault et al., 2015; Gladkov et al., 2017; Forró et al., 2018; Pigareva et al., 2021; Winter-Hjelm et al., 2023). These technical issues may be resolved by utilizing more sophisticated geometries, e.g., PDMS devices bearing nanometer-scale channels that only allow the penetration of dendritic spines (Mateus et al., 2022). Another challenge is the exponential increase in computational demand required to generate the state-transition matrix $\mathbf{T}$ and conduct its eigendecomposition as the number of nodes $N$ increases. While further refinement of theoretical methodology remains essential for handling larger networks, the current results underscore that by appropriately coarse-graining the system, the model provides a robust mathematical framework for analytically understanding the structure-function relationships in complex networks.

*5. Conclusions*

By combining in vitro experiments with in silico modeling, we investigated the functional role of directional connections in a network of cortical neurons and showed that beyond generating a flow of information, it enriches the network dynamics by reducing excessive synchrony in cultured neuronal networks and increasing the integration-segregation balance. In silico models using SNN and state-transition models provided non-trivial predictions that (1) a cooperative relationship between modularity and directionality is critical, and (2) the configuration employed in the in vitro experiment is quasi-optimal to maximize the integration-segregation balance. This study demonstrates the critical role that engineered neuronal networks grown in microfluidic devices can play in bridging theoretical models with biology, particularly in exploring the relationships between structure and function in neuronal networks. Moreover, this study advances the development of brain-on-a-chip technology as a method to reconstitute neuronal networks that more accurately mimic the nervous system, which is beneficial for pharmacological research in drug discovery (Courte et al., 2023; Sakaibara et al., 2024) and for biocomputing research (Yada et al., 2021; Sumi et al., 2023; Cai et al., 2023).

**Declaration of competing interest**

The authors declare that they have no competing financial interests or personal relationships that may have influenced the work reported in this study.

**Data availability**

Data will be made available on request.

**Acknowledgements**

We thank Mr. Iori Morita at the Fundamental Technology Center, Research Institute of Electrical Communication (RIEC), Tohoku University for the fabrication of the photomasks. The work was partly supported by MEXT Grant-in-Aid for Transformative Research Areas (A) and (B) “Multicellular Neurobiocomputing” (21H05164, 24H02330, 24H02332), JSPS KAKENHI (22H03657, 22K19821, 22KK0177, 23H00251, 23H02805, 23H03489, 23K11259, 23K16958), JST-CREST (JPMJCR19K3), the WISE Program for AI Electronics by Tohoku University, and the Cooperative Research Project Program of the Research Institute of Electrical Communication (RIEC) at Tohoku University. N.F. was supported by JST Moonshot R&D Program Grant Number JPMJMS2021. This research was partly carried out at the Laboratory for Nanoelectronics and Spintronics, RIEC, Tohoku University.

Supplementary Materials for

# Directional intermodular coupling enriches functional complexity in biological neuronal networks

Nobuaki Monma, Hideaki Yamamoto*, Naoya Fujiwara, Hakuba Murota, Satoshi Moriya, Ayumi Hirano-Iwata, and Shigeo Sato

***Correspondence:** hideaki.yamamoto.e3@tohoku.ac.jp (Hideaki Yamamoto)

**This file includes:**

## Extended methods

*Functional modularity*

Following previous work (Newman, 2004), functional modularity $Q$ was calculated from correlation coefficient $r_{ij}$, given by:

$$Q = \frac{1}{2M}\sum_{ij}\left(r_{ij} - \frac{k_i k_j}{2M}\right)\delta(m_i, m_j),$$

where $M$ represents the sum of all correlation coefficients $\sum_{i,j}^{N} r_{ij}$, $k_i$ denotes the sum of the coefficients in the $i$-th row, $\delta$ is the Kronecker delta, and $m_i$ is an index of the module neuron $i$ belongs to.

*Axonal model*

To provide a better illustration of the spiking neural network (SNN) model (Fig. 4A), the axonal elongation was numerically simulated following the previous work (Orlandi et al., 2013). Initially, the boundary was established by arranging 16 squares, each with sides measuring 100 μm, in a 4 × 4 configuration. All squares were connected hierarchically with 20 rectangles, each with dimensions of 100 μm by 7 μm. Seven neurons were randomly placed in each square, and each neuron generated an axon by extending 1-μm segments with an angle $\Delta\theta$ relative to the previous segment. The probability distribution of $\Delta\theta$ was given by:

$$p(\Delta\theta) = \frac{1}{\sqrt{2\pi\sigma_\theta}} \exp\left(\frac{(\Delta\theta)^2}{2\sigma_\theta^2}\right),$$

where $\sigma_\theta$ is the standard deviation set to 5°. If the segment extended beyond the boundary, $\Delta\theta$ was calibrated by determining the minimal angle needed to guide the segment back to the boundary, with a resolution of 1°. The extension process was repeated until the total length of all segments reached 1100 μm.

**Supplementary Figures**

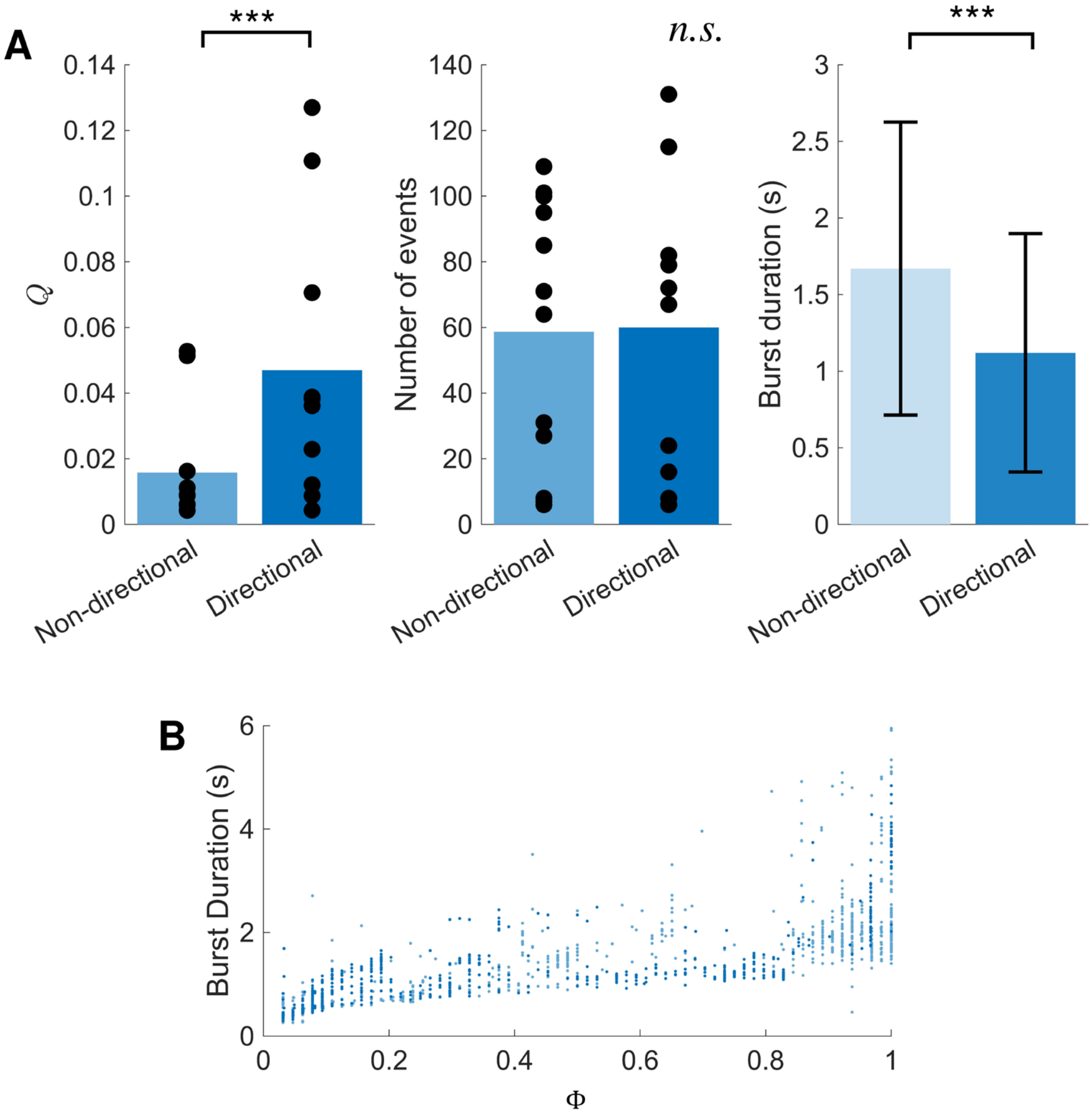

Fig. S1. Statics of spontaneous activity in non-directional ($n$ = 12) and directional ($n$ = 10) networks at 10 div. (A) Left: Functional modularity $Q$. Middle: Number of collective activity events. Right: Durations of collective activity events. (*p < 0.05; **p < 0.01; ***p < 0.001, n.s., no significance, two-tailed Student's $t$-test.) (B) The relationship of global network activation (GNA) Φ and event durations.

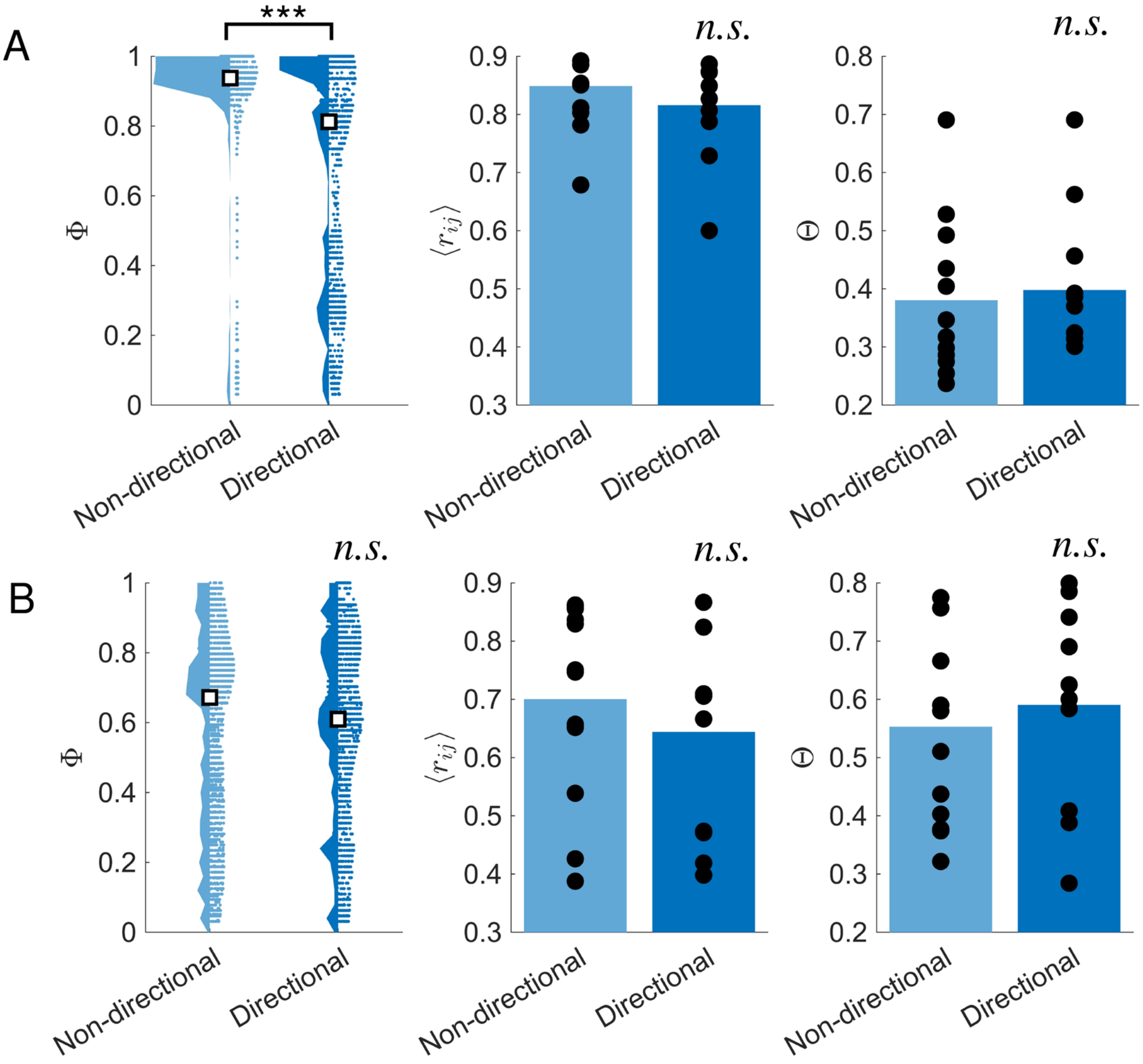

Fig. S2. Dependence of global network activation (GNA) Φ, mean correlation $\langle r_{ij} \rangle$, and functional complexity Θ on cultivation days. The statics were derived from the spontaneous activity in (A) 14 div and (B) 21 div. The number of samples were $n$ = 12 and 10 for the non-directional and directional networks, respectively. (*p < 0.05; **p < 0.01; ***p < 0.001, n.s., no significance, two-tailed Student's $t$-test.)

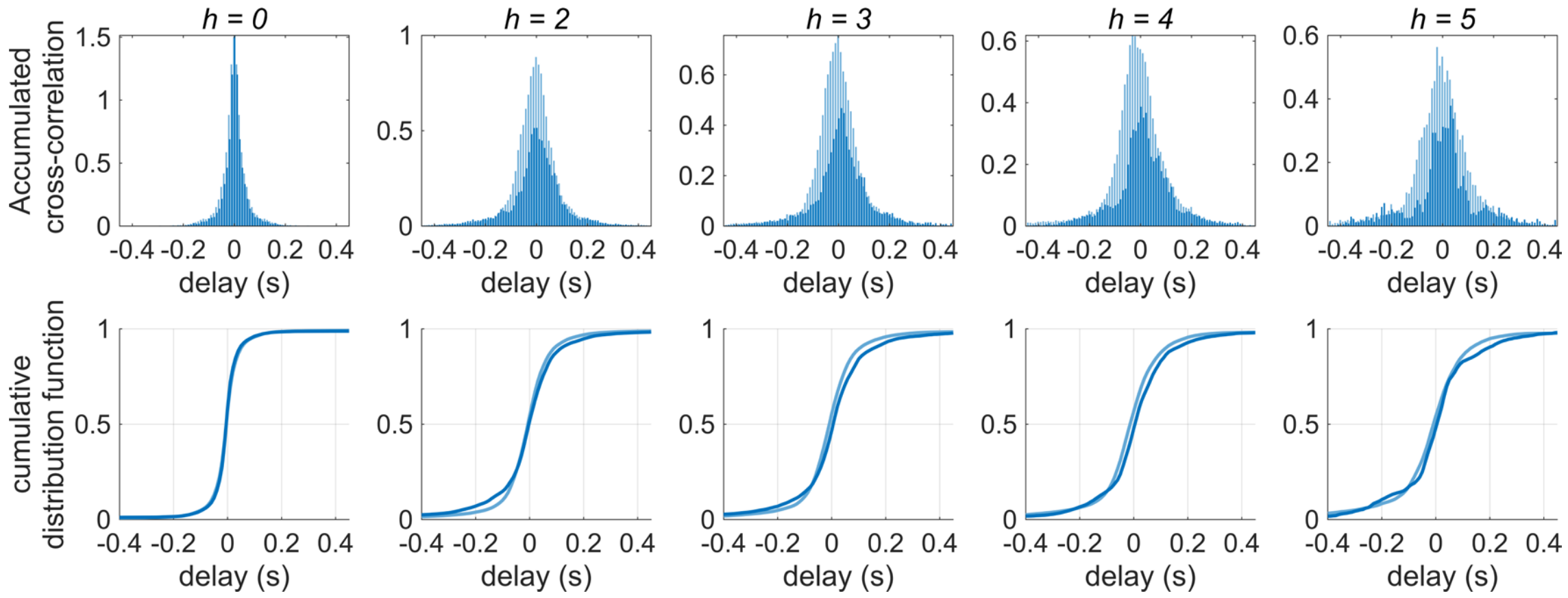

Fig. S3. Cross-correlation function analysis in directional (dark blue) and non-directional networks (light blue). Data for $h = 1$ and 6 are presented in the main text. Top: Accumulated cross-correlation function. Bottom: Cumulative distribution. The positive and negative delays represent the forward and reverse propagations, respectively.

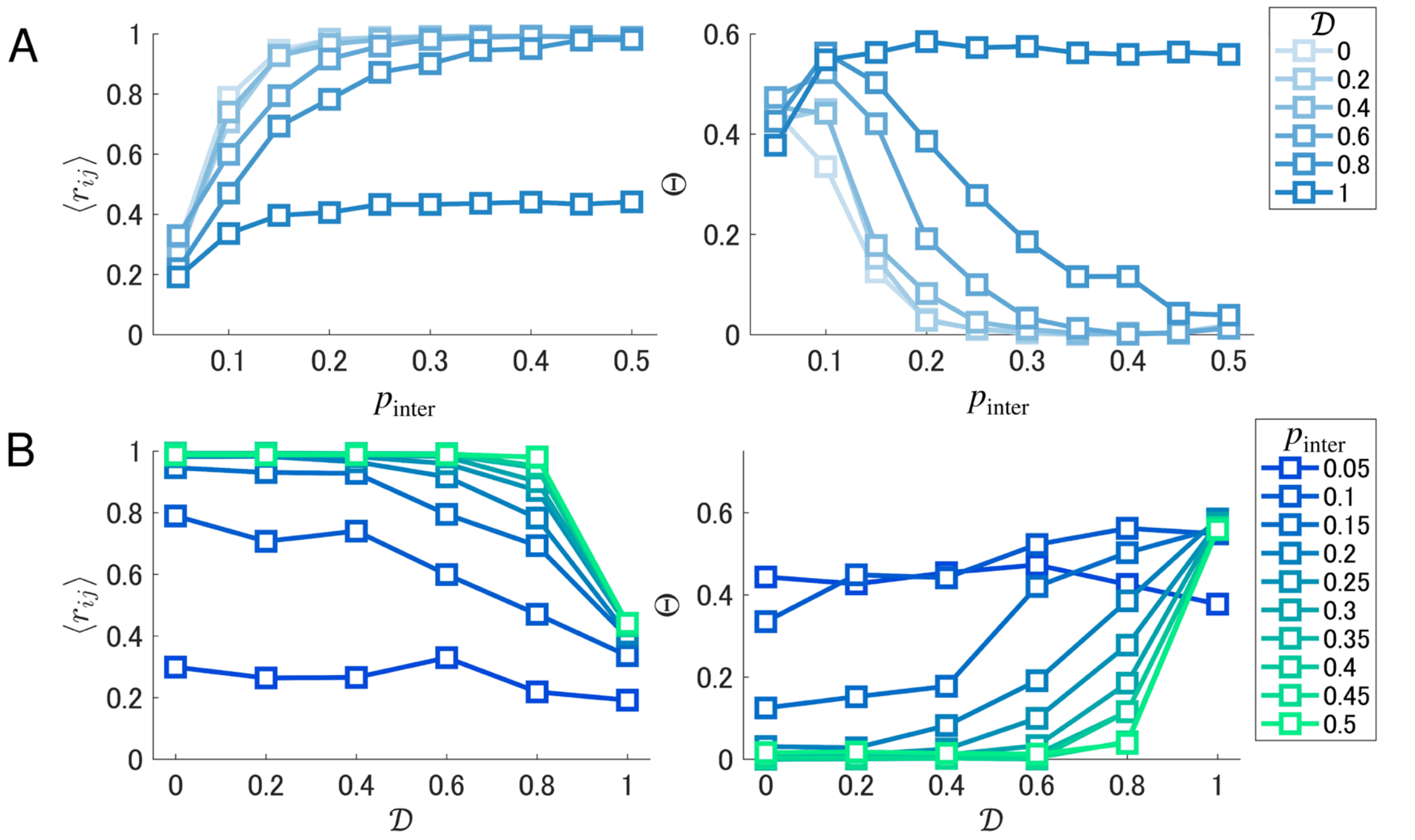

Fig. S4. Mean correlation and functional complexity across the entire parameter space of (A) $p_{\text{inter}}$ and (B) $\mathcal{D}$ in SNN simulations. Each square plot represents the mean of 40 network realizations.

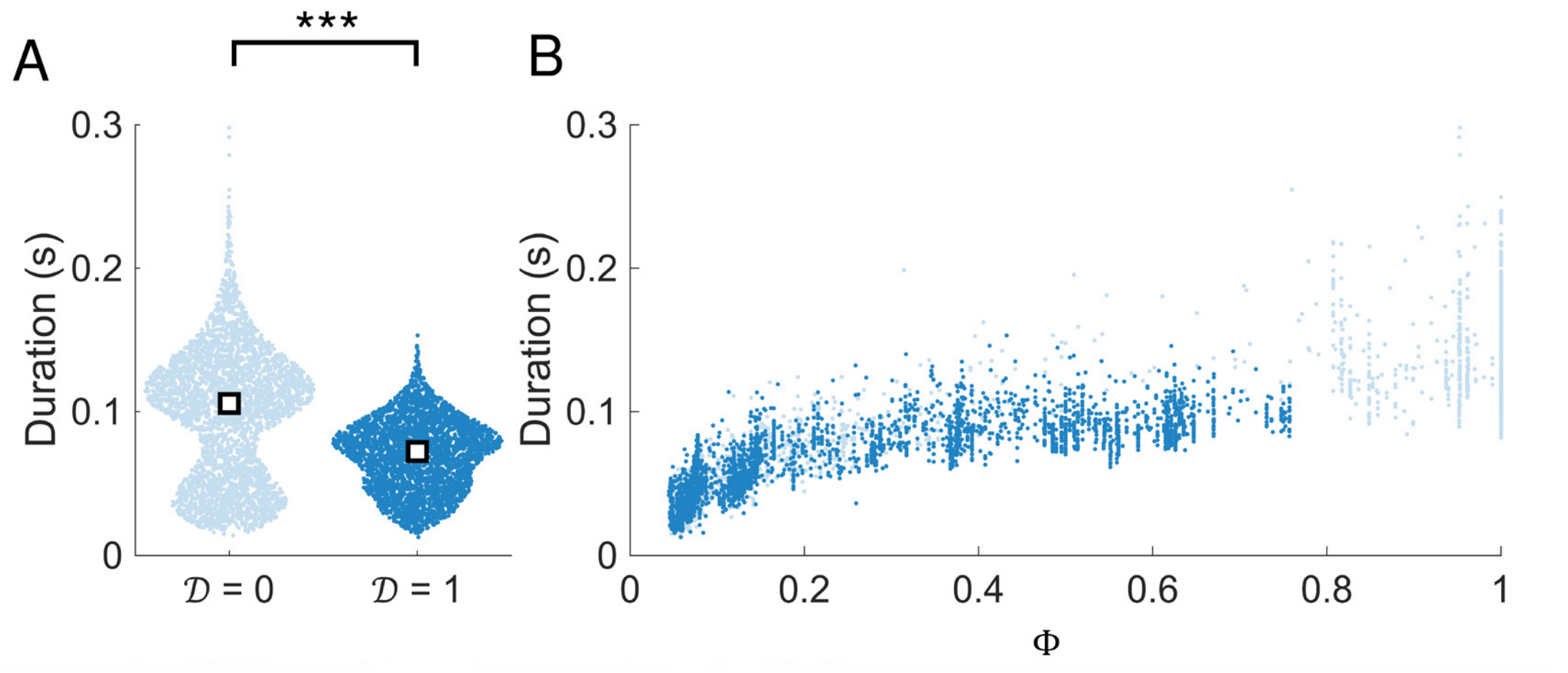

Fig. S5. Duration of collective activity events in non-directional ($\mathcal{D}$ = 0) and fully-directional ($\mathcal{D}$ = 1) networks in SNN simulations. (A) The distribution of durations. Each square represents median of 40 network realizations. (***p < 0.001, two-tailed Student's *t*-test). (B) The relationship between GNA and event duration.

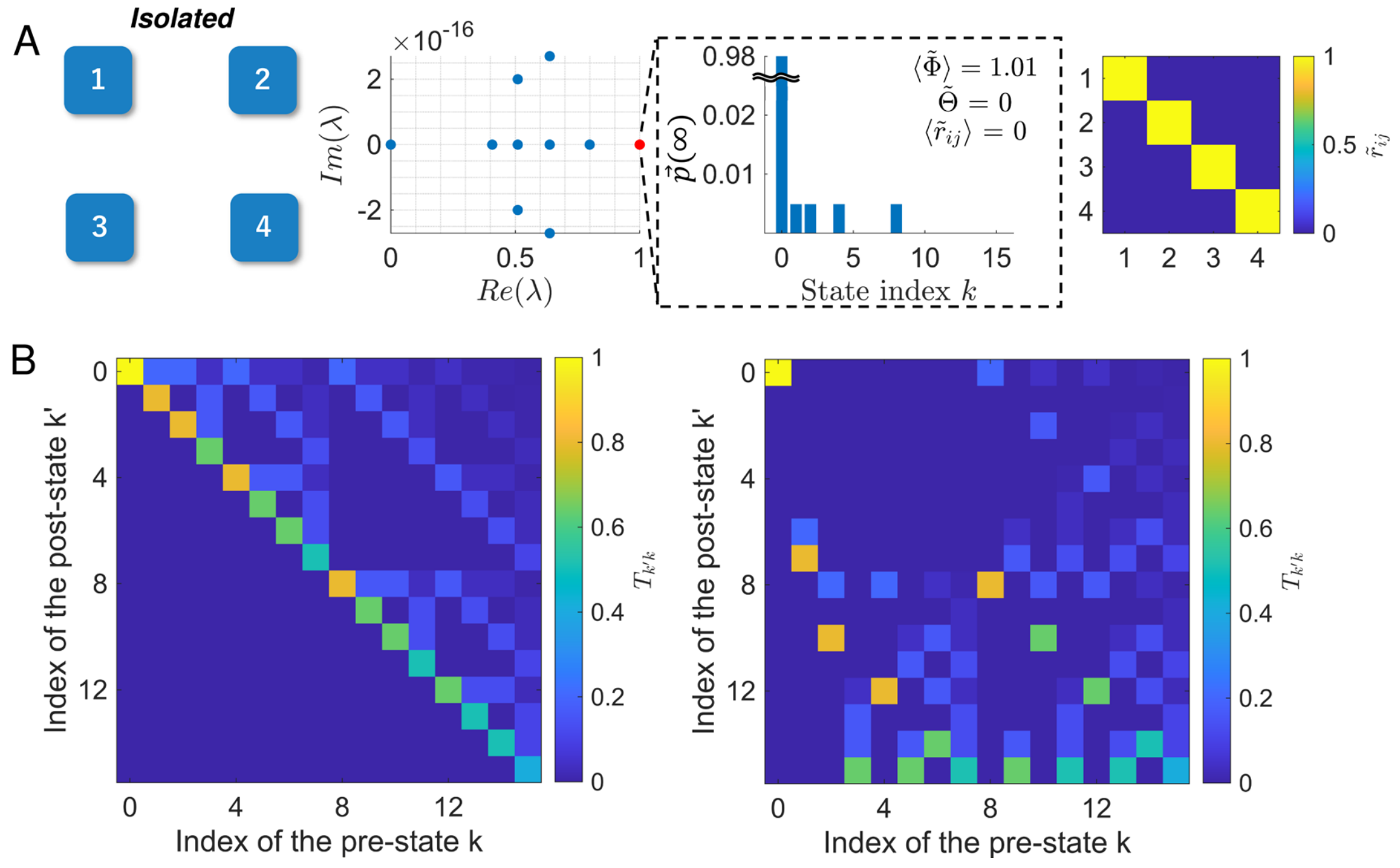

Fig. S6. Representative analysis of a 4-node network. (A) Left: Structure of the *isolated* network. Middle: Eigenspectrum obtained from the state-transition matrix **T** and the probability of observing network states $\vec{p}(\infty)$. The latter was derived from the eigenvector corresponding to the eigenvalue equal to 1. Right: Correlation matrix $\tilde{r}_{ij}$. (B) State-transition matrix **T** of the *isolated* (left) and *feedforward* networks (right) with $\sigma = 10^{-3}$ and $\gamma = 0.8$.

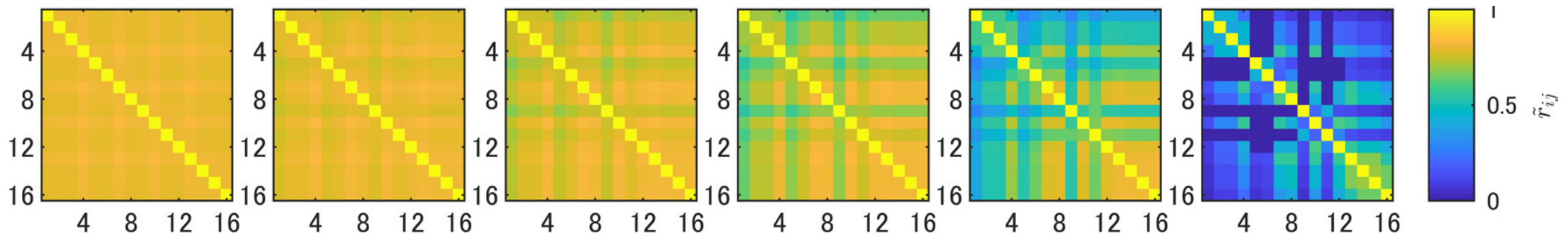

Fig. S7. Dependence of correlation matrix $\tilde{r}_{ij}$ on the directionality $\widetilde{\mathcal{D}}$ in state-transition model with the topology depicted in Fig. 5C. Correlation matrices are displayed from the left to right in ascending order of directionality $\widetilde{\mathcal{D}}$ = {0, 0.2, 0.4, 0.6, 0.8, 1}.

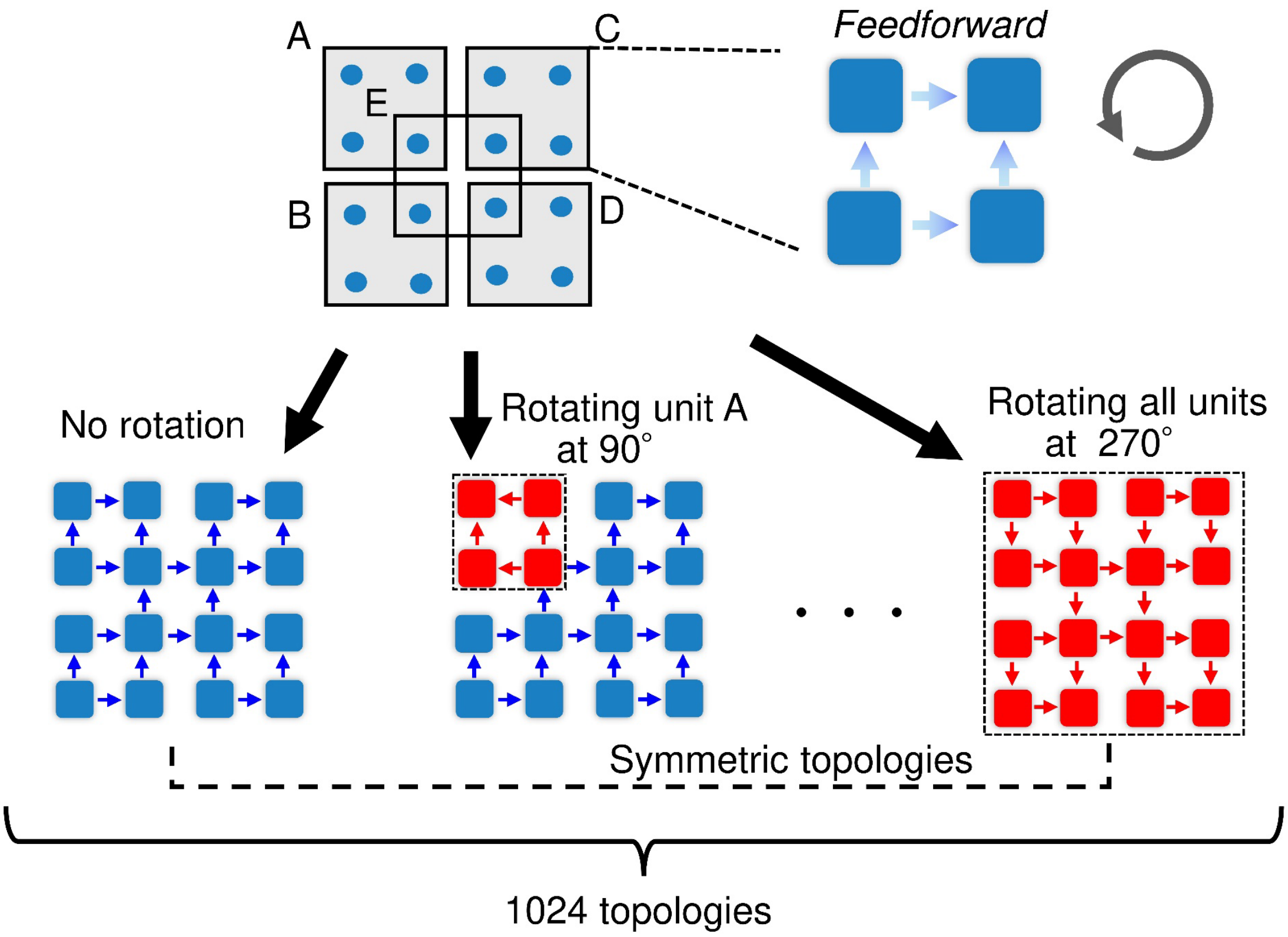

Fig. S8. Methodology to derive topologies explored in Section 3.3.4. These topologies were generated by rotating 2×2-node groups connected in a feedforward manner. The sample size was $4^5$ (=1,024) as each set of five units can be rotated 0, 90°, 180°, or 270° counterclockwise. In certain cases, different operations generate topologies that are symmetric, such as the one without any rotation (bottom left) and the one obtained by rotating all units by 270° (bottom right).

## Supplementary Tables

Table 1. The pseudo-code for generating state-transition matrix $\mathbf{T} = [T]_{k'k}$.

---

**Algorithm 1** Generating State-Transition Matrix $\mathbf{T} = [T]_{k'k}$

---

**Input** Connection matrix $W$, firing states $\vec{s_k}, \vec{s_{k'}}$

**Output** Probability of transition from $\vec{s_k}$ to $\vec{s_{k'}}$, denoted as $T_{k'k}$

**Definition**

$$A(f_k(i)) = \begin{cases} \sigma & \text{if } f_k(i) = 0, \\ f_k(i) & \text{if } 0 < f_k(i) < 1, \\ 1 & \text{if } 1 \leq f_k(i), \end{cases} \qquad \triangleright \text{ Activation function}$$

$$P(s_{k'}(i) \mid s_k(i)) = \begin{cases} 1 - A(f_k(i)) & \text{if } (s_{k'}(i), s_k(i)) = (0,0), \\ 1 - \gamma & \text{if } (s_{k'}(i), s_k(i)) = (0,1), \\ A(f_k(i)) & \text{if } (s_{k'}(i), s_k(i)) = (1,0), \\ \gamma & \text{if } (s_{k'}(i), s_k(i)) = (1,1), \end{cases} \qquad \triangleright \text{ Conditional probability}$$

**procedure** GenerateProbability($W$, $\vec{s_k}$, $\vec{s_{k'}}$) ▷ Main process

Calculate the synaptic input vector as:

$$\vec{f_k} \Leftarrow \mathbf{W}^\top \vec{s}_k$$

Using $\vec{f_k}$, calculate the transition probability from $\vec{s_k}$ to $\vec{s_{k'}}$ as:

$$T_{k'k} \Leftarrow \prod_{i=1}^{N} P(s_{k'}(i) | s_k(i))$$

**return** $T_{k'k}$

**end procedure**

---

Table 2. Overview of notations

| Notation | Description |
|---|---|
| | **Shared notations** |
| $N$ | Number of neurons |
| $t$ | Time |
| $\Phi$ | Global network activation (GNA) |
| $r_{ij}$ | Correlation coefficient between neurons $i$ and $j$ |
| $\Theta$ | Functional complexity |
| $B$ | Number of bins (= 20) |
| $\mathcal{D}$ | Degree of directionality ([0,1]) |
| | **In vitro experiment** |
| $F(t)$ | Fluorescence intensity at time $t$ |
| $x_i(t)$ | Firing rate of neuron $i$ |
| $S_i(t)$ | Binary spike train of neuron $i$, with 1 for a firing onset |
| $R_{ij}(\tau)$ | Cross-correlation function with time delay $\tau$ |
| $T$ | Duration of a recording (= 1200 s) |
| $\Delta t$ | Recording time step (= 0.01 s) |
| $\delta(\cdot,\cdot)$ | Kronecker delta |
| $h$ | Number of hops ([0, 6]) |
| $M$ | Normalization constant of mean correlation |
| $m_i$ | Index of a module with neuron $i$ ([0,16]) |
| $p(r_{ij})$ | Probability distribution of $r_{ij}$ |
| | **Spiking neural network** |
| $v$ | Membrane potential |
| $u$ | Recovery variable |
| $a, b, c, d$ | Izhikevich parameters ($a = 0.02, b = 0.2, c = -65, d = 8$) |
| $\eta$ | White Gaussian noise (mean = 0, SD = 1.0) |
| $I_i$ | Synaptic current injected to neuron $i$ |
| $\mathrm{d}t$ | Simulation time step (= 0.1 ms) |
| $\mathbf{W}$ | Connection matrix (mean = 3.0, SD = 1.0) |
| $p_{\mathrm{inter}}$ | Probability of inter-modular connection ([0, 0.5]) |
| $\tau$ | Synaptic time constant (= 10 ms) |
| $H(\cdot)$ | Binary step function |
| $\Delta_{ij}$ | Synaptic delay (mean = 0.6, SD = 0.1) |
| | **State-transition model** |
| $k$ | Index of the firing state |
| $\vec{s}_k$ | Firing state assigned index $k$ |
| $\mathbf{W}$ | Connection matrix |
| $\vec{f}_k$ | Synaptic input in firing state $\vec{s}_k$ |
| $A(\cdot)$ | Activation function |
| $\sigma$ | Probability of spontaneous activation (= $10^{-3}$ or $10^{-12}$) |
| $\gamma$ | Probability of persistent activation (= 0.8) |
| $\mathbf{T}$ | State-transition matrix |
| $\lambda_i$ | $i$-th eigenvalue of $\mathbf{T}$ |
| $\vec{x}_i$ | Eigenvector associated with eigenvalue $\lambda_i$ |
| $\vec{p}(\infty)$ | Probability vector at equilibrium ($t \to \infty$) |
| $\sim$ | Symbol representing the state-transition model |